\documentclass[conference]{IEEEtran}

\IEEEoverridecommandlockouts

\usepackage{graphicx}
\usepackage{subfigure}
\usepackage{amsfonts}
\usepackage{amssymb}
\usepackage{amsmath}
\usepackage{amsthm}
\newtheorem{mydef}{Definition}
\newtheorem{mythm}{Theorem}
\usepackage{mdwlist}
\usepackage{changepage}
\usepackage{makecell,rotating,multirow,diagbox}
\usepackage{epstopdf}
\usepackage{color}
\usepackage{cite}
\usepackage{textcomp}
\usepackage{bm}
\usepackage{url}

\usepackage{threeparttable}
\usepackage[usestackEOL]{stackengine}  
\usepackage[shortlabels]{enumitem}

\usepackage{booktabs}
\usepackage{framed}
\usepackage{multicol}
\usepackage{tikz}
\usepackage{algorithm}
\usepackage{algorithmic}
\usepackage{caption}

\makeatletter
\newif\if@restonecol
\makeatother

\usepackage[algo2e,linesnumbered,ruled,vlined]{algorithm2e}
\usepackage{amsmath}

\usepackage[justification=centering]{caption}
\usetikzlibrary{positioning}
\tikzset{>=stealth}

\def\BibTeX{{\rm B\kern-.05em{\sc i\kern-.025em b}\kern-.08em
		T\kern-.1667em\lower.7ex\hbox{E}\kern-.125emX}}

\begin{document}

	\title{Open-Pub: A Transparent yet Privacy-Preserving Academic Publication System based on Blockchain
	}

	\author{Yan Zhou, Zhiguo Wan and Zhangshuang Guan
		\thanks{Corresponding author: Zhiguo Wan.}
		\thanks{Y. Zhou, Z. Wan and Z. Guan are with the School of Computer Science and Technology, Shandong University,
			Qingdao, 266237, China (e-mail:\{yanzhousdu@mail.,wanzhiguo@,guanzs@mail.\}sdu.edu.cn).}
		\thanks{This work is supported by National Natural Science Foundation of China undert Grant No. 61972229.}
	}

	\maketitle

\begin{abstract}
	Academic publications of latest research results are crucial to advance the development of all disciplines.
	However, there are several severe disadvantages in current academic publication systems.
	The first is the misconduct during the publication process due to the opaque paper review process.
	An anonymous reviewer may give biased comments to a paper without being noticed because the comments are seldom published for evaluation.
	Second, access to research papers is restricted to only subscribers, and even the authors cannot access their own papers.

	To address the above problems, we propose Open-Pub, a decentralized, transparent yet privacy-preserving academic publication scheme using the blockchain technology.
	In Open-Pub, we first design a threshold identity-based group signature (TIBGS) that protects identities of
	signers using verifiable secret sharing.
	Then we develop a strong double-blind mechanism to protect the identities of authors and reviewers.
	With this strong double-blind mechanism, authors can choose to submit papers anonymously, and validators distribute papers anonymously to reviewers on the blockchain according to their research interests.
	This process is publicly recorded and traceable on the blockchain so as to realize transparent peer preview.
	To evaluate its efficiency, we implement Open-Pub based on Ethereum and conduct comprehensive experiments to evaluate its performance, including computation cost and processing delay.
	The experiment results show that Open-Pub is highly efficient in computation and processing anonymous transactions.
\end{abstract}


\section{Introduction}
\label{sec:introduction}
In academia, publishing the latest research achievements on academic publications can significantly promote advances of sciences and technologies.
In current publication systems, the publication procedure roughly includes paper submission, assignment, review and the final publication.
Most mainstream academic publishers work in this way, such as Elsevier, Springer, Institute of Electrical and Electronics Engineers (IEEE) and Association for Computing Machinery (ACM). Meanwhile, there are some online systems like EDAS\footnote{https://edas.info/} and Easychair\footnote{https://easychair.org/} to manage submitted manuscripts for conferences and journals.

Current publication systems have several problems, which should be addressed for the benefit of the whole community.
The first problem is academic misconducts due to the opaque reviewing process.
Normally, a paper is given to several reviewers to examine its contributions and novelties objectively, and reviewers' comments usually are not open to the public.
Reviewers are likely to provide comments that are not solely based on research merits for some reasons, e.g. personal preferences, conflict of interest and intense competitions among researchers.

The current review process also lacks mechanisms to motivate reviewers to provide constructive and unbiased comments.
A reviewer is seldom rewarded for his/her valuable comments, making it difficult to motivate reviewers to give constructive comments.

Besides, it is important to share the latest research achievements in academia.
Currently, many preprint systems publish and share research results in different disciplines without peer review, e.g. arXiv\footnote{https://arXiv.org/}, bioRxiv\footnote{https://eprint.iacr.org/} and IACR eprint\footnote{https://bioRxiv.org/}.
Preprint systems still suffer from problems due to centralization, including lack of transparency and misconducts.
Moreover, research results on these preprint systems may be problematic because most of them are not reviewed by peers.

In the current paper review process, many journals and conferences adopt the double-blind or single-blind approach.
For the double-blind review, the reviewers and the authors do not know each other, while in the single-blind review, the reviewers know the authors.
Authors have to disclose their identities to a centralized entity, and the centralized entity knows the authorship of every submitted manuscript.
Therefore, effective protection of the anonymity of authors can make the review process more complete.

The emerging blockchain~\cite{Bitcoin} technology can be utilized to solve the problems in current academic publication systems.
The decentralization, transparency and immutability of blockchain can improve the transparency and fairness of the publishing process.
The blockchain technology is originally designed as an open, distributed ledger without any trusted party.
Due to its advantages of decentralization, transparency, fault-tolerance and credibility, the blockchain has been applied in many fields such as finance, insurance, notarization, healthcare, internet of things and social networks.

Obviously, making the whole publishing process transparent in the blockchain can greatly promote fair review and academic sharing.
But if the identities of the authors and the reviewers are disclosed, this will also cause the reviewer's comments to be affected by personal emotions.
Maintaining anonymity while keeping the process open and transparent is a challenge in itself.
Moreover, the identities of the authors are hidden before the review and are made public after the review, which makes it more difficult to hide the identities in the blockchain.

In this paper, we propose a transparent and privacy-preserving decentralized academic publication system named Open-Pub, which is based on a consortium blockchain operated by multiple validators.
The validators can be served by existing publishers or government agencies to maintain the system.
Everyone can trace the entire process from the submission of the paper to the final publication.
To achieve anonymity during the review process of Open-Pub, we develop the threshold group signature scheme TIBGS from the identity-based group signature in \cite{smart2009identity}.
Reviewers are rewarded according to the quality of their comments, and hence they are motivated to provide authors with unbiased and constructive comments.

To achieve a fair publishing system, we use the blockchain to make the whole review process publicly visible and design a strong double-blind mechanism.
While keeping the system transparent, the identities of authors and reviewers can be well hidden during the review process.
Eventually, their identities will be made public, promoting transparency throughout the process. Public scrutiny and double-blind review together provide a guarantee of fair peer review.

The contributions of this paper can be summarized as follows:

\begin{itemize}	
	\item We propose Open-Pub, a transparent and privacy-preserving academic publication system that is based on the blockchain technology.
	To manage keys for Open-Pub and develop a strong double-blind mechanism, we also designed TIBGS, a Threshold Identity-based Group Signature scheme.
	To the best of our knowledge, this is the first decentralized privacy-preserving academic publication system based on blockchain.
	\item We design decentralized account management based on the threshold signature\cite{stathakopoulous2017threshold}, which is used to manage public assets.
	\item We formulate a security model for Open-Pub and prove its security by giving a simulation-based proof. We also provide discussion and analysis on security, performance and further enhancements for Open-Pub.
	\item We implement Open-Pub by modifying Ethereum \cite{wood2014ethereum} source code and conduct comprehensive experiments to evaluate its performance. We test the computation and communication costs for each type of operations in Open-Pub, and the result shows that Open-Pub is efficient in both computation and communication.
\end{itemize}

The remainder of the paper is structured as follows.
We first review research work related to blockchain privacy protection and application of blockchain in the academic publication in Section II.
Then we provide preliminaries on our proposal, including cryptographic building blocks in Section III.
Next, we describe a threshold identity-based group signature algorithm TIBGS and analyze its security in Section IV.
We then present Open-Pub in detail in Section V. After that, we give a comprehensive discussion and analysis of Open-Pub in Section VI.
We describe details on the implementation of Open-Pub and evaluate its performance in Section VII.
Finally, concluding remarks are given in Section VIII.
\section{Related Work}
In this section, we introduce the work related to privacy-preserving blockchain and the application of blockchain in academic publishing.

\subsection{Privacy-Preserving Blockchains}
The first blockchain system Bitcoin~\cite{Bitcoin} was invented by Satoshi Nakamoto in 2008.
In public blockchains, all transactions are public and can be verified by every participant.
Transaction amounts and the links between transactions are publicly visible. However, privacy issues emerge as a serious problem for blockchains. As a result, a number of privacy-preserving solutions for blockchains have been proposed recently. 

Monero \cite{van2013cryptonote} was a successful privacy-preserving cryptocurrency using ring signature \cite{rivest2001leak}. The ring signature allows a member of a set to sign on behalf of the set. Unlike the group signature, there is no way to revoke the anonymity of a ring signature. Although the ring signature provides strong anonymity, there are some limitations in efficiency and security.
First, the size of a ring signature is directly proportional to the number of participants.
Second, its transactions (especially RingCT transactions) are very large in size, with almost thousands of bytes per transaction, which adds storage space for the entire blockchain record. Monero is an untraceable digital currency, with transaction details completely invisible to the public.

Another popular privacy-preserving cryptocurrency is Zerocash \cite{sasson2014zerocash}, an anonymous cryptocurrency built from Bitcoin.
Zerocash makes use of zk-SNARKs (zero-knowledge succinct non-interactive arguments of knowledge) \cite{ben2013snarks} proofs to hide transaction amounts and participants. Zerocash provides strong anonymity and transaction privacy protection for the blockchain, but it is computationally expensive in generating transaction proofs. In addition, the zk-SNARKs algorithm requires a trusted setup step. If the adversary is aware of the secret randomness used in the setup, the adversary can generate deceptive proofs for false statements, and the false statements are indistinguishable from true statements.

Privacy-preserving blockchains like Monero or Zerocash can protect privacy, but they cannot be used directly for academic publishing. The group signature can reveal identity, which is an important feature of paper review. We improved the group signature to make it more suitable for blockchain and academic publishing.

\subsection{Blockchain for Publication}
Many studies utilize the blockchain technology to promote scientific publication.
Novotny \emph{et al.}~\cite{novotny2018permissioned} highlight the transparency of the blockchain system for academic publishing.
Janowicz \emph{et al.}~\cite{janowicz2018prospects} present an outline that aims to combine distributed ledger technologies and academic publishing.
Leible \emph{et al.}~\cite{leible2019review} introduce the adaptability, challenges and research potential between blockchain and open science.
Heaven \emph{et al.}~\cite{heaven2019bitcoin} introduce the advantages and challenges of applying blockchain to scientific publishing.
Duh \emph{et al.}~\cite{duh2019publish} present some social dilemmas occurring in academic publishing under a strategic game setting and show that building a trusted scientific community is the key to promote a publish-and-flourish culture.
Mohan \emph{et al.}~\cite{mohan2019use} emphasize the use of blockchain to tackle academic misconducts.

Eureka~\cite{schaufelbuhl2019eureka,niya2019blockchain} is a blockchain-based scientific publishing platform, aiming to address problems in the current academic publishing industry, such as traditional inefficient processes, long delays, and lack of fair financial incentives. Eureka maps the review process to the blockchain through smart contracts and designs a token-based incentive mechanism.

Orvium~\cite{orvium2019} focuses on integrating the blockchain technology into the publication lifecycle. The platform aims to reduce the cost of publishing and access, create better incentives for peer reviewers, increase transparency in the peer review process, and promote better sharing of research data.

PubChain \cite{wang2019pubchain} uses blockchain, smart contracts and the peer-to-peer file-sharing system IPFS to implement a decentralized open-access publication platform. PubChain utilizes the blockchain technology to incentivize participation of authors, readers and reviewers and carries out a simulation to study the proposed decentralized scoring system.

Mackey \emph{et al.}~\cite{mackey2019framework} propose a governance framework for scientific publishing, aiming to enhance transparency, accountability, and trust in the publishing process. The ultimate goal of the framework is to create an ecosystem allowing participants to eventual self-govern and agree on how to to enforce the rules and norms fairly.

Coelho \emph{et al.}~\cite{coelho2019decentralising} propose a system to solve incentive problems of traditional systems in science communication and publishing, and present a minimal working model to define roles, processes, and expected results of the novel system.

Tenorio-Fornes \emph{et al.}~\cite{tenorio2019towards} propose a decentralized publication system for open science based on blockchain and IPFS, and develop a proof-of-concept prototype.
In addition to fairness and transparency, the authors also noticed the privacy requirement.

Unfortunately, all these studies did not solve the privacy problem during the peer review process, while Open-Pub aims to tackle this challenge for blockchain-based academic publication.
\section{Preliminaries}
In this section, we will introduce some cryptographic techniques used in Open-Pub, including Verifiable Secret Sharing (VSS), an asymmetric encryption algorithm and two signature algorithms.

\begin{table}
	\vspace{-0.2cm}
	\center
	\caption{Notations}
	\vspace{-0.2cm}
	\label{table}
	\setlength{\tabcolsep}{3pt}
	\begin{tabular}{|p{70pt}|p{165pt}|}
		\hline
		Notation  &  Meaning \\
		\hline
		$\lambda$ & security parameter\\
		$(\mathsf{mpk, msk_i})$ & master public key and master private key share \\
		$\mathsf{grpID}$ & the group identifier\\
		$\mathbf S$ & a set of group managers\\
		$(\mathsf{gsk_i, gvk_i})$ & group secret key share and group verify key share\\
		$\mathtt{userID}$ &the user identifier\\
		$\mathtt{usk}$ & group private key\\
		$\mathsf{\sigma}$ & signature\\
		$\mathsf(k, n)$ & threshold\\
		\hline
	\end{tabular}
	\label{tab1}
	\vspace{-0.2cm}
\end{table}


\subsection{Pedersen's Verifiable Secret Sharing}

A ($k,n$) Pedersen's verifiable secret sharing scheme~\cite{pedersen1991non} enables $n$ participants to share a random value $x$ without a trusted third party, and at least $k$ participants ($1 \leq k \leq n$) can participants to restore $x$. Before $x$ is restored, the random secret value $x$ is kept secret from all participants. Each participant obtains a share ${x_i}$ known by participant $i$ only. More importantly, each participant can verify the validity of ${x_i}$ to detect invalid messages sent by malicious participants.
Let ${P_1}$, ${P_2}$, $\cdots$, ${P_n}$ be the $n$ participants. The protocol for ${P_i}$ is:
\begin{enumerate}
	\setlength{\leftskip}{-8pt}
	\item Choose a random number $s_{i,0} \in Z_q$;
	\item Distribute $s_{i,0}$ verifiably among ${P_1}$, ${P_2}$, $\cdots$, ${P_n}$ and ${P_j}$ can get $s_{i,j}$;
	\item Verify $n-1$ received shares;
	\item After receiving $n-1$ correct shares, ${P_i}$ compute the share $s_i = {s_{1,i}} + {s_{2,i}} + {s_{3,i}} + \cdots + {s_{n,i}}$. The complete secret $s = s_{1,0} + s_{2,0} + \cdots+ s_{n,0}$ is shared among $n$ participants.
\end{enumerate}

Later we will use Pedersen's VSS scheme in our protocol, and use $(k,n)$-VSS to denote a Pedersen's VSS scheme for $(k,n)$ secret sharing.

\subsection{Cryptographic Building Blocks}
An asymmetric encryption \cite{bellare1994optimal} scheme can be represented by a tuple of polynomial-time algorithms $\Pi_{enc} = \bf{(Setup, KeyGen, Enc, Dec)}$.
\begin{itemize}
	\item{${\bf{Setup}}(\mathsf1^{\lambda}) \rightarrow \mathsf{pp}$}. On input a security parameter ${\lambda}$, this algorithm generates a list of public parameters $\mathsf{pp}$.
	
	\item{${\bf{KeyGen}}(\mathsf{pp}) \rightarrow \mathsf{(pk, sk)}$}. On input a list of public parameters $\mathsf{pp}$, this algorithm generates a public/secret key pair $(\mathsf{pk},\mathsf{sk})$.
	
	\item{${\bf{Enc}}({m},\mathsf{pk}) \rightarrow c$}. With a public key $\mathsf{pk}$, this algorithm encrypts an input plaintext $m$ to output a ciphertext $c$.
	
	\item{${\bf{Dec}}(c, \mathsf{sk}) \rightarrow m$}. With a secret key $\mathsf{sk}$, this algorithm decrypts an input ciphertext $c$ to output a plaintext $m$.
\end{itemize}

A signature scheme can be represented by a tuple of polynomial-time algorithms $\Pi_s$ = $\bf{(Setup}$, $\bf{KeyGen}$, $\bf{Sign}$, $\bf{Verify)}$.
\begin{itemize}
	\item{${\bf{Setup}}(\mathsf1^{\lambda}) \rightarrow \mathsf{pp}$}. On input a security parameter ${\lambda}$, this algorithm generates a list of public parameters $\mathsf{pp}$.
	
	\item{${\bf{KeyGen}}(\mathsf{pp}) \rightarrow (\mathsf{pk},\mathsf{sk}$)}. On input a list of public parameters $\mathsf{pp}$, this algorithm generates a public/private key pair $(\mathsf{pk},\mathsf{sk})$.
	
	\item{${\bf{Sign}}({m},\mathsf{sk}) \rightarrow \mathsf{sig}$}. With a private key $\mathsf{sk}$, this algorithm generate a signature $\mathsf{sig}$ corresponding to the message $m$.
	
	\item{${\bf{Verify}}(\mathsf{pk},{m},\mathsf{sig}) \rightarrow \mathsf\{{0,1}\}$}.This algorithm can verify whether the signature $\mathsf{sig}$ is generated by private key $\mathsf{sk}$ corresponding to public key $\mathsf{pk}$.
\end{itemize}

A ($k, n$) threshold signature \cite{boneh2001short} on a message $m$ is a single, constant-sized aggregate signature that passes verification if and only if at least $k$ out of the $n$ participants sign $m$.
Note that the verifier does not need to know the identities of the $k$ signers.
A ($k,n$)  threshold signature scheme can be represented by a tuple of polynomial-time algorithms $\Pi_{ts} = \bf{(Setup, ThresKeyGen, ThresSign,}$ $\bf{SigShareVer, SigShareComb, Verify)}$.
\begin{itemize}
	\item{${\bf{Setup}}(\mathsf1^{\lambda}) \rightarrow \mathsf{pp}$}. On input a security parameter ${\lambda}$, this algorithm generates a list of public parameters $\mathsf{pp}$.
	
	\item{${\bf{ThresKeyGen}}(\mathsf{pp}, k, n) \rightarrow (\mathsf{PK}, \mathsf{sk}_i, \mathsf{vk}_i$)}. On input a list of public parameters $\mathsf{pp}$, this algorithm generates a public key $\mathsf{PK}$, a set of $n$ secret key shares $\{ \mathsf{sk_1, sk_2, ..., sk_n}\}$ and a set of verification keys $\{ \mathsf{vk_1, vk_2, ..., vk_n}\}$.
	
	\item{${\bf{ThresSign}}({m},\mathsf{sk}_i) \rightarrow \mathsf{sig}_i$}. Each participant signs the message $m$ with a secret key share $\mathsf{sk}_i$ and output a signature share $\mathsf{sig}_i$.
	
	\item{${\bf{SigShareVer}}(\mathsf{PK}, \mathsf{vk}_i, {m},\mathsf{sig}_i) \rightarrow \mathsf\{{0,1}\}$}. The algorithm can verify the correctness of the signature share $\mathsf{sig}_i$ by $\mathsf{PK}$ and the corresponding $\mathsf{vk}_i$.
	
	\item{${\bf{SigShareComb}}(\mathsf{{sig}_i}s, k) \rightarrow \mathsf{sig}$}. With at least $k$ valid signature shares $\mathsf{sig}_i$'s, this algorithm calculates the complete signature $\mathsf{sig}$.
	
	\item{${\bf{Verify}}(\mathsf{PK}, {m},\mathsf{sig}) \rightarrow \mathsf\{{0,1}\}$}. The algorithm can verify the correctness of the complete signature $\mathsf{sig}$ by $\mathsf{PK}$.
	
\end{itemize}

$\Pi_{enc}$ used in our scheme needs to satisfy key indistinguishability and ciphertext indistinguishability under chosen-ciphertext attack \cite{rackoff1991non}. $\Pi_{s}$ and $\Pi_{ts}$ should satisfy unforgeability and robustness against adaptive identity \cite{bellare2001key} and chosen message attack \cite{bellare1998relations}.
\section{TIBGS: Threshold Identity-based Group Signature}

In this section, we describe TIBGS, a threshold identity-based group signature algorithm.
In accordance with TIBGS, Open-Pub can manage identity-based keys and carry out cryptographic operations in a decentralized way.
The group signature \cite{camenisch1997efficient} scheme allows a member of a group to sign a message anonymously without leaking identity information. A group manager can open the group signature to disclose the true identity of the signer.

TIBGS involves $3$ different participants: $n$ group managers, group users and verifier.
In contrast to ordinary group signature schemes, the number of group managers has increased from $1$ to $n$.

In TIBGS, we use ($k,n$)-VSS to decentralize the master private key $\mathsf {msk}$, the group secret key $\mathsf {gsk}$ and the group verify key $\mathsf {gvk}$ to $n$ managers, with each manager holding only a secret shadow.
Therefore, each manager can only produce a portion of the group private key $\mathsf{usk_i}$ for the group user through $\mathsf{gsk}$ and $\mathsf{userID}$.
With $k$ $\mathsf{usk_i}$'s from $k$ group managers, a user can compute complete group private key $\mathsf{usk}$, which is used to sign anonymously on behalf of the group.
From a group signature $\mathsf{\sigma}$ from the group, others can not find the signer for this signature and a verifier can use $\mathsf{grpID}$ and master public key $\mathsf {mpk}$ to verify the correctness of the signature.
Finally, the identities of anonymous signers can be exposed by at least $k$ group managers.
As a result, TIBGS realizes decentralize cryptographic operations, including key generation, signature and opening.

\begin{itemize}
	\item{${\bf{Setup}}(\mathsf1^{\lambda},k,n) \rightarrow (\mathsf{mpk},\mathsf{msk}_i)$}. Each manager can run this algorithm to generate master public key $\mathsf{mpk}$ and master private key share $\mathsf{msk}_i$.
	
	\item{${\bf{GrpSetUp}}(\mathsf{grpID}, i, \mathsf{msk}_i,k,n) \rightarrow (\mathsf{gsk}_i,\mathsf{gvk}_i$)}. This algorithm on input of $\mathsf{grpID}$ and $\mathsf{msk}_i$ and outputs a group secret/verify key pair ($\mathsf{gsk}_i,\mathsf{gvk}_i$) corresponding to $i$th manager.
	\item{${\bf{ExtShare}}(\mathsf{userID},\mathsf{gsk}_i) \rightarrow \mathsf{usk}_i$}. The group manager executes the algorithm and outputs the group private key share $\mathsf{usk}_i$, which is sent to the user.
	\item{${\bf{ReconstKey}}(\mathsf{userID},\{\mathsf{usk}_i\}_{i\in \mathbf{S}},\{\mathsf {gvk}_{i}\}_{i\in \mathbf{S}}) \rightarrow \mathsf{usk}$}. The user executes the algorithm to reconstruct its full private key from the secret shares obtained from managers.
	\item{${\bf{Sign}}(m,\mathsf{usk}) \rightarrow \mathsf{\sigma}$}. Each user can execute the algorithm and generate a signature $\mathsf\sigma$ corresponding to the message $m$.
	\item{${\bf{Verify}}\mathsf(m,\sigma,\mathsf{mpk,grpID}) \rightarrow \mathsf\{{0,1}\}$}. This algorithm can verify whether the signature is generated by user in the group $\mathsf{grpID}$.
	\item{${\bf{OpenPart}}(\mathsf{gsk}_i,\sigma,m) \rightarrow \mathsf{ok}_i$}. The group manager can execute the algorithm and obtain an intermediate result $\mathsf{ok}_i$.
	\item{${\bf{Open}}(k,\{\mathsf{ok}_i\}_{i\in \mathbf{S}}) \rightarrow \mathsf{userID}$}. The group manager can execute the algorithm and reveal the identifier $\mathsf{userID}$ of the user who produced the signature $\mathsf\sigma$ corresponding to the message $m$.
\end{itemize}

We formulate the security of TIBGS with the full-anonymity experiment and the full-traceability experiment in Appendix.B.

\begin{mydef}[Full-anonymity]
	Let $\Pi = (\bf{Setup}$, $\bf{GrpSetUp}$, $\bf{ExtShare}$, $\bf{ReconstKey}$, $\bf{Sign}$,  $\bf{Verify}$, $\bf{OpenPart}$, $\bf{Open})$ be a threshold identity-based group signature scheme.
	We say that $\Pi$ is fully anonymous if for all sufficiently large security parameter $k\in \mathbb{N}$ and any proper
	probabilistic polynomial time (PPT) adversary $\mathcal{A}$, its advantage
	$\mathsf{Adv}_{\Pi, \mathcal{A}}^{\mathsf{anon}}(1^\lambda)$ = $|\mathsf{Pr}[{\bf Exp}_{\Pi, \mathcal{A}}^{\mathsf{anon}}(1^\lambda)=1]$ - $\frac{1}{2}|$ is negligible.
\end{mydef}

\begin{mydef}[Full-traceability]
	Let $\Pi = (\bf{Setup}$, $\bf{GrpSetUp}$, $\bf{ExtShare}$, $\bf{ReconstKey}$, $\bf{Sign}$,  $\bf{Verify}$, $\bf{OpenPart}$, $\bf{Open})$ be a threshold  identity-based group signature scheme.
	We say that $\Pi$ is fully traceable if for all sufficiently large security parameter $k\in \mathbb{N}$ and any proper
	probabilistic polynomial time (PPT) adversary $\mathcal{A}$, its advantage
	$\mathsf{Adv}_{\Pi, \mathcal{A}}^{\mathsf{trace}}(1^\lambda)$ = $\mathsf{Pr}[{\bf Exp}_{\Pi, \mathcal{A}}^{\mathsf{trace}}(1^\lambda)=1]$ is negligible.
\end{mydef}

For a threshold scheme, the following robustness property is defined as follows:
\begin{mydef}[Robustness]
	A TIBGS scheme is said to be robust if it computes a correct output even in the presence
	of a malicious attacker that makes the corrupted managers deviate from the normal execution.
\end{mydef}

\begin{mythm} Assuming that the IBGS scheme in~\cite{smart2009identity} is fully anonymous, the above TIBGS scheme is also fully anonymous. (The proof is given in the Appendix.B)
\end{mythm}

\begin{mythm} Assuming that the IBGS scheme in~\cite{smart2009identity} is fully traceable, the above TIBGS scheme is also fully traceable.
\end{mythm}
The proof is similar to that of the full-anonymity theorem. Due to limited space, the proof is omitted.

The following theorem states the robustness of the proposed TIBGS scheme:
\begin{mythm}
	Assuming that $n\geq 2k-1$ where $(k,n)$ is the threshold of the proposed TIBGS scheme, then it is
	robust in the presence of up to $k-1$ corrupted managers.
\end{mythm}

\begin{proof}
	It is easy to see that $k$ honest group managers are required to generate a valid group private key $\mathsf{usk}$, and at most $k-1$ managers can be corrupted. In addition, each group manager obtains a group verification key $\mathsf{gvk}_i$ corresponding to its secret group key share $\mathsf{gsk}_i$. The group verification keys are published for verifying group private key shares $\mathsf{usk}_i$'s.
	As a result, the user can check the validity of the secret key shares $\mathsf{usk}_i$ using $\mathsf{gvk}_i$, and reconstruct his key from $k$ valid user key shares. To sum up, the proposed TIBGS scheme is robust if $n\geq 2k-1$.
\end{proof}
\section{Open-Pub: The Transparent yet Privacy-Preserving Academic Publication System}

In this section, we first provide the system model of Open-Pub and then present the threat model.
After that, we describe the design of Open-Pub, including its transaction management and threshold identity management.

\subsection{System Model}

\begin{figure}[!ht]
	\vspace{-0.2cm}
	\center
	\includegraphics[width=0.5\textwidth]{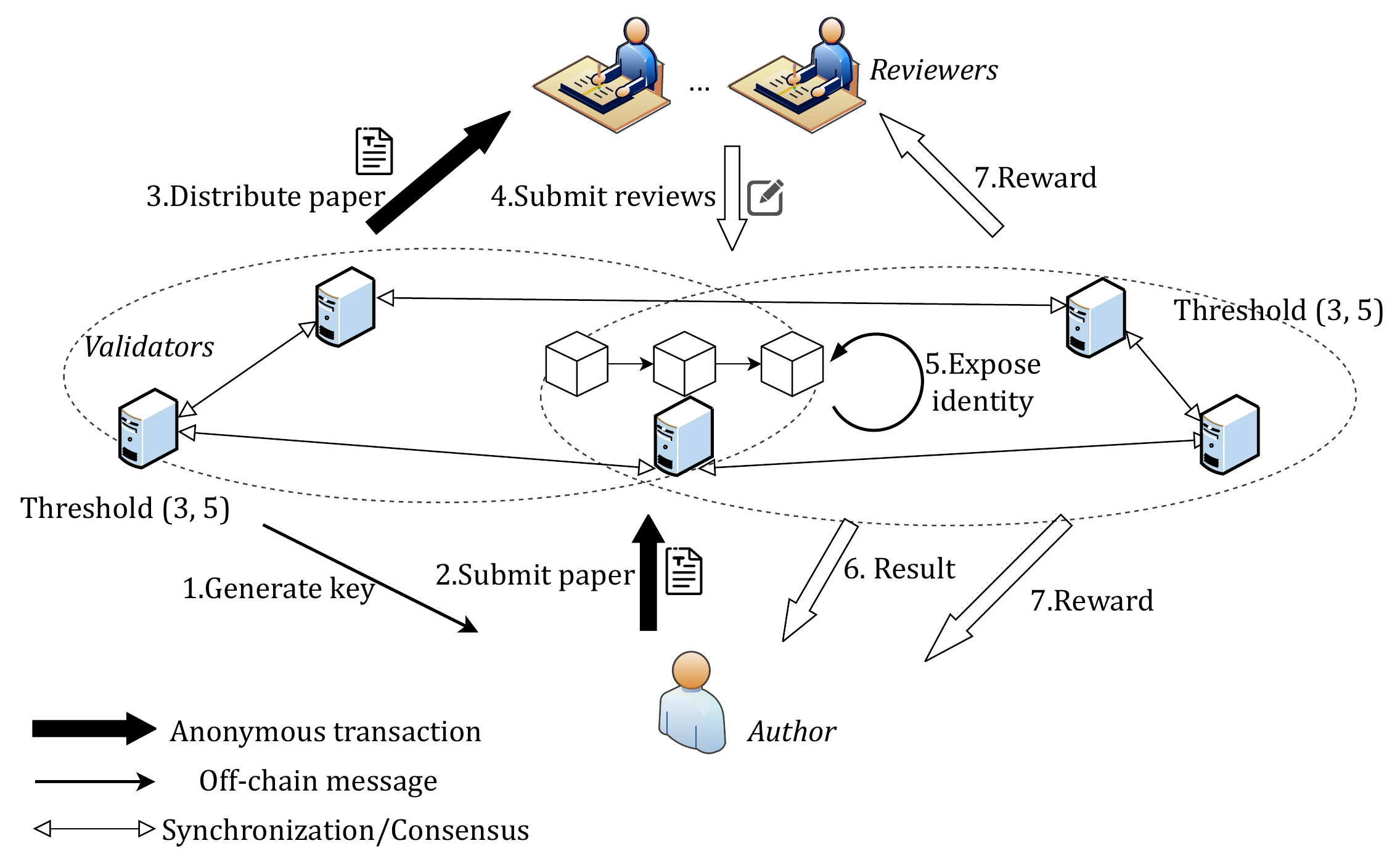}
	\vspace{-0.2cm}
	\caption{The system architecture and workflow of Open-Pub with 5 validators and threshold 3.} \label{fig:system}
	\vspace{-0.2cm}
\end{figure}

The system architecture of Open-Pub and the workflow are depicted in Fig.~\ref{fig:system}.
The Open-Pub system involves $4$ different participants with the following roles and responsibilities:
\begin{itemize}
	\setlength{\leftskip}{-10pt}
	\item {\bf Validator.} Validators validate transactions and broadcast transactions in Open-Pub. Besides, validators are responsible for the distribution of group private keys, papers and rewards. After the paper review, validators collaborate to reveal the anonymous author and make the final result public.
	Finally, validators send rewards to authors and reviewers. All validators maintain Open-Pub to work properly.

	\item {\bf Author.} In the system, author needs to request group private key shares $\mathsf{usk}_i$'s from validators and computes complete group private key $\mathsf {usk}$. With $\mathsf {usk}$, author can submit papers anonymously. During the review process, the author does not know who the reviewer is until receiving reviewer comments.
	If the paper is accepted, the author will receive a reward from the validators.

	\item {\bf Reviewer.} Before the review process, reviewers will register in the blockchain based on their research areas, which helps to find suitable reviewers. After receiving a paper from the validator, the reviewer puts forward his own review comments and scores, which are conducted in the form of sending transactions. The reviewer cannot know the real identity of the anonymous author until the validators reveal the author. After the review process, reviewers will receive review fees.

	\item {\bf Reader.} Readers can comment and score on papers or comments, but readers’ comments do not affect the final results of the papers.
\end{itemize}

On registration, an author can get a group private key share $\mathsf {usk}_i$ from each validator and calculate the complete the group private key $\mathsf {usk}$ with a sufficient number of shares (Step 1).
Authors can submit papers anonymously, and anonymous submissions are signed using the group private key $\mathsf {usk}$ (Step 2).

After the validators receive the submitted papers, they randomly select reviewers for the papers based on the research area (Step 3).
To hide the identities of reviewers, the validators will use the public keys of the reviewers to encrypt the identities and paper information, and then send the ciphertext and deadline to the blockchain.
Only the corresponding reviewer can use his/her private key to decrypt the ciphertext to get the paper information.
Reviewers review and grade papers using their real names, and they do not know the real identity of the anonymous authors in the process (Step 4).

When the deadline is reached, the validators will cooperate to expose the anonymous author, and publish the result of the paper based on existing reviews (Step 5-6).
To motivate reviewers and authors, validators will pay review fees to reward reviewers, and if the papers are accepted, the validators will pay rewards to encourage the authors (Step 7).
The blockchain makes the entire review process visible to all members, including submission, distribution, review, opening and reward, making the review process open and transparent.

We emphasize that how much the authors and the reviewers should be rewarded is an independent research problem,
and we leave it as future work. These rewards are created by validators following a pre-determined rule, like
the incentive mechanism in Bitcoin. Any appropriate rewarding mechanism can be used in Open-Pub, but we assume
these rewards are fixed in the following description.


\subsection{Threat Model}
We assume a Byzantine threat model in which the adversary can compromise no more than 1/3 validators of blockchain.
Instead of following the specified protocol, the compromised validators will act arbitrarily and may collude with each other to coordinate attacks, including injecting, modifying and dropping messages during participating in the protocol. However, the adversary is assumed to have limited computation resources and cannot break the cryptosystem used in our proposal.

Open-Pub aims to achieve the following design goals:
\begin{itemize} 
	\setlength{\leftskip}{-10pt}

	\item {\bf Accountability.} Whenever there is misconduct or abuse in the system, the system should be able to identify the author or the reviewer according to the corresponding transaction.

	\item {\bf Anonymity.} The identity of the anonymous author is kept secret during the review process, and multiple validators collaborate to reveal the identity of the anonymous author after the review is complete.
	During the review process, no one except the distributor and the reviewers themselves can know which papers the reviewers have been assigned.

	\item{\bf Recoverability.} When the group private key of an author is lost, it should be recoverable.
\end{itemize}
We formulate the security of Open-Pub with the full-anonymity experiment in Appendix.C.

\subsection{Open-Pub}
We will describe in detail how Open-Pub utilizes the TIBGS to implement a decentralized privacy-preserving academic publication system on the blockchain.

Based on TIBGS and asymmetric encryption, we develop a strong double-blind mechanism to protect the identities of authors and reviewers in Open-Pub.
Authors obtain the group private key $\mathsf {usk}$ from validators and use $\mathsf {usk}$ to anonymously submit papers to blockchain.
The verifier receives the paper and encrypts the paper with the public key of reviewer.
Then, the verifier publicly transmits the ciphertext to the public account to avoid revealing the identity of the reviewer.
Only the corresponding reviewer can decrypt the corresponding ciphertext to obtain the paper information.
The anonymity of authors is guaranteed by group signature, and the anonymity of reviewers is guaranteed by asymmetric encryption.

The Open-Pub system consists of the following $7$ steps: system initialization, registration, submission, distribution, review, open and reward.
To implement these steps, we create five types of transactions including $\mathsf{tx_{transfer}}$, $\mathsf{tx_{submit}}$, $\mathsf{tx_{distribute}}$, $\mathsf{tx_{review}}$ and $\mathsf{tx_{open}}$, and we introduce the processing logic of these transactions.
We specify that the first of the transaction structure is the public key of the sender $\mathsf{pk_{sender}}$ and the second is the public key of the receiver $\mathsf{pk_{receiver}}$.
We stipulate that $\mathsf{sig}$ represents the ordinary signature, $\mathsf{gsig}$ represents the group signature and $\mathsf{tsig}$ represents the threshold signature.
Details of these algorithms are given in Fig.~\ref{fig:alg}.

\begin{figure*}[!ht]
	\vspace{-0.5cm}
	\begin{framed} \scriptsize
		\vspace{-0.4cm}
		\raggedright
		\begin{multicols}{2}
			
			$\underline{\bf{SystemInitialization}}$\\
			\vspace{1pt}
			\begin{itemize}
				\setlength{\leftskip}{-6pt}
				\item inputs: $\lambda, k, n, \mathsf{grpID}$
				\item outputs: $\mathsf{mpk}, \mathsf{msk}_i, \mathsf{gsk}_i, \mathsf{gvk}_i, \mathsf{pk},\mathsf{sk}, \mathsf{acc_{pub}}, \mathsf{tsk}_i, \mathsf{tvk}_i$
			\end{itemize}
			\begin{enumerate}
				\setlength{\leftskip}{-6pt}   
				\item $\mathsf{mpk}, \mathsf{msk}_i =$ $\bf{TIBGS.Setup}$ $(\mathsf1^{\lambda},k,n)$;       
				\item $\mathsf{gsk}_i,\mathsf{gvk}_i$ = $\bf{TIBGS.GrpSetUp}$$(\mathsf{grpID}, i, \mathsf{msk}_i,k,n)$;
				\item $\mathsf{pp}_s = \Pi_{s}.\bf{Setup}(\mathsf1^{\lambda})$;
				\item $\mathsf{pk, sk} = {\Pi_{s}.\bf{KeyGen} (\mathsf{pp}_s)}$;
				\item $\mathsf{pp}_{ts} = \Pi_{ts}.\bf{Setup}(\mathsf1^{\lambda})$;
				\item $\mathsf{acc_{pub}},\mathsf{tsk}_i, \mathsf{tvk}_i$ = $\Pi_{ts}.\bf{ThresKeyGen}$$(\mathsf{pp}_{ts}, k, n)$;
				\item \textbf{return} {$\mathsf{mpk}, \mathsf{msk}_i, \mathsf{gsk}_i, \mathsf{gvk}_i, \mathsf{pk},\mathsf{sk}, \mathsf{acc_{pub}}, \mathsf{tsk}_i, \mathsf{tvk}_i$};
			\end{enumerate} 
			$\underline{\bf Registration}$ \\
			\begin{itemize}
				\setlength{\leftskip}{-6pt}
				\item inputs: $\lambda, \mathsf{type}, \mathsf{userID}, \mathsf{acc_{pub}}, \mathsf{gsk}_i, \mathsf {gvk}_{i}, \$_\mathsf{deposit}$, $k$
				\item outputs: $\mathsf{pk, sk}, \mathsf {tx_{transfer}}$, $\mathsf{usk}$
			\end{itemize}
			\begin{enumerate}
				\setlength{\leftskip}{-6pt}
				\item $\mathsf{pp_{s}} = \Pi_{s}.\bf{Setup}(\mathsf1^{\lambda})$;
				\item $\mathsf{pk, sk} = {\Pi_{s}.\bf{KeyGen} (\mathsf{pp})}$;
				\item If {$\mathsf{type} = 0 || \mathsf{type} = 1 || \mathsf{\$_\mathsf{deposit} < GetBalance(pk)}$} /*$\mathsf{GetBalance()}$ is a function to get the account balance*/
				\begin{enumerate}
					\setlength{\leftskip}{-6pt}
					\item \textbf{return} $\mathsf{pk, sk}$;
				\end{enumerate}
				\item $\mathsf v =  \$_\mathsf{deposit}$;
				\item $\mathsf {tx_{origin}} = (\mathsf{pk, acc_{pub}, userID, v})$;
				\item $\mathsf{sig} = {\Pi_{s}.\bf{Sign}(\mathsf {tx_{origin}, sk})}$;
				\item $\mathsf {tx_{transfer}} = (\mathsf{tx_{origin}, sig})$;
				\item set the balance of $\mathsf{pk}$ to  $\mathsf{GetBalance(pk)-v}$;
				\item set the balance of $\mathsf{acc_{pub}}$ to  $\mathsf{GetBalance(acc_{pub})+v}$;
				\item send $\mathsf{userID}$ to each validator to get $\mathsf{usk}_i $=$ \bf{TIBGS}.{ExtShare}(\mathsf{userID},$ $\mathsf{gsk}_i)$ and store them in set $\mathsf{S_{{usk}}}_i$;
				\item If $\mathsf{GetNum(S_{{usk}})}_i \geq k$   /*$\mathsf{GetNum()}$ is a function that gets the size of a set*/
				\begin{enumerate}
					\setlength{\leftskip}{-6pt}
					\item $\mathsf{usk}=\bf{TIBGS.ReconstKey}$ $(\mathsf{userID},\{\mathsf{usk}_i\}_{i\in \mathbf{S}},$ $\{\mathsf {gvk}_{i}\}_{i\in \mathbf{S}})$;
					\item \textbf{return} $\mathsf{pk, sk}, \mathsf{tx_{transfer}}$, $\mathsf{usk}$; 
				\end{enumerate}
			\end{enumerate}
			$\underline{\bf Submission}$ \\
			\begin{itemize}
				\setlength{\leftskip}{-6pt}
				\item inputs: $\mathsf{field, paper, pk_{annoymity}, acc_{pub}, usk}$
				\item outputs: $\mathsf{tx_{submit}}$
			\end{itemize}
			
			\begin{enumerate}
				\setlength{\leftskip}{-6pt}
				\item $\mathsf {tx_{origin}} = \mathsf{\mathsf {(pk_{annoymity}, acc_{pub}, field, paper)}}$;
				\item $\mathsf{gsig} = {\bf{TIBGS.Sign}(\mathsf {tx_{origin}, usk})}$;
				\item \textbf{return} $\mathsf {tx_{submit}} = (\mathsf{tx_{origin}, gsig})$;
			\end{enumerate}
			$\underline{\bf Distribution}$ \\
			\begin{itemize}
				\setlength{\leftskip}{-6pt}
				\item inputs: $\mathsf{h_{tx_{submit}}, reviewerIDs, endtime, pk, sk,  acc_{pub}}$		
				\item outputs: $\mathsf{tx_{distribute}}$			
			\end{itemize}
			
			\begin{enumerate}
				\setlength{\leftskip}{-6pt}
				\item For each $\mathsf{reviewerID}$ in $\mathsf{reviewerIDs}$
				\begin{enumerate}
					\setlength{\leftskip}{-6pt}
					\item $\mathsf{pk_r = GetPK(reviewerID)}$/*$\mathsf{GetPK()}$ is a function to get the $\mathsf{pk}$ of the $\mathsf{ID}$*/;
					\item $\mathsf{r = GenRandom()}$/*$\mathsf{GenRandom()}$ is a function to generate a new random number */;
					\item store $\mathsf{c} = \Pi_{enc}.{\bf{Enc}}\mathsf{(( h_{tx_{submit}}, reviewerID, r), pk_r)}$ in set $\mathsf{S_{ciphertext}}$;
				\end{enumerate}	
				\item $\mathsf {tx_{origin}} = (\mathsf{\mathsf {pk, acc_{pub}, h_{tx_{submit}}, S_{ciphertext}, endtime}})$;
				\item $\mathsf{sig} = {\Pi_{s}.\bf{Sign}(\mathsf {tx_{origin}, sk})}$;
				\item \textbf{return} $\mathsf {tx_{distribute}} = (\mathsf{tx_{origin}, sig})$;
			\end{enumerate}
			$\underline{\bf Review}$ \\
			\vspace{1pt}
			\begin{itemize}
				\setlength{\leftskip}{-6pt}
				\item inputs: $\mathsf{{reviewerID, tx_{distribute}, pk, sk, acc_{pub}}}$		
				\item outputs: $\mathsf{tx_{review}}$
			\end{itemize}
			\begin{enumerate}
				\setlength{\leftskip}{-6pt}
				\item For each $\mathsf{c}$ in $\mathsf{tx_{distribute}.S_{ciphertext}}$
				\begin{enumerate}
					\setlength{\leftskip}{-6pt}
					\item If $\mathsf{h_{tx_{submit}}, reviewerID, r} = \Pi_{enc}.{\bf{Dec}}\mathsf{(c, sk)}$
					\begin{enumerate}
						\setlength{\leftskip}{-6pt}
						\item break;
					\end{enumerate}
				\end{enumerate}
				\item find the paper through $\mathsf{h_{tx_{submit}}}$, and review it to get $\mathsf{comment}$ and $\mathsf{score}$;
				\item $\mathsf {tx_{origin}} = ({\mathsf {pk, acc_{pub}, h_{tx_{submit}}, reviewerID, r, comment, score}})$;
				\item $\mathsf{sig} = {\Pi_{s}.\bf{Sign}(\mathsf {tx_{origin}, sk})}$;
				\item \textbf{return} $\mathsf {tx_{review}} = (\mathsf{tx_{origin}, sig})$;	
			\end{enumerate}
			$\underline{\bf Open}$ \\
			\vspace{1pt}
			\begin{itemize}
				\setlength{\leftskip}{-6pt}
				\item inputs: $\mathsf{tx_{submit}, tx_{distribute}, tx_{review}}$'s $, \mathsf{result}, \mathsf{gsk}_i, k, \mathsf{pk, sk,}$ $\mathsf{acc_{pub}}$
				\item outputs: $\mathsf{tx_{open}}$
			\end{itemize}
			\begin{enumerate}
				\setlength{\leftskip}{-6pt}
				\item If {$\mathsf{Time() \geq tx_{distribute}.endtime}$/*$\mathsf{Time()}$ is a function to get the current time*/}
				\begin{enumerate}
					\setlength{\leftskip}{-6pt}
					\item $\mathsf{{ok}_i} = \bf{TIBGS.OpenPart}(\mathsf{gsk}_i, \mathsf{tx_{submit}.gsig}, \mathsf{tx_{submit}})$;
					\item send $\mathsf{tx_{submit}}$ to other validator to get other $\mathsf{ok}_i$'s and store them in set $\mathsf{S_{ok}}_i$;
					\item If {$\mathsf{GetNum(S_{ok}}_i) \geq k$}
					\begin{enumerate}
						\setlength{\leftskip}{-6pt}
						\item $\mathsf{userID} = {\bf{TIBGS.Open}} (k,\{\mathsf{ok}_i\}_{i\in \mathbf{S}})$;
					\end{enumerate}
					\item $\mathsf{reviewerIDs = tx_{review}s.reviewerID}$;
					\item $\mathsf{tx_{origin}} = (\mathsf{pk, acc_{pub}, \mathsf{h_{tx_{submit}}}, userID, result, reviewerIDs}$ $)$;
					\item $\mathsf{sig} = {\Pi_{s}.\bf{Sign}(\mathsf {tx_{origin}, sk})}$;
					\item \textbf{return} $\mathsf {tx_{open}} = (\mathsf{tx_{origin}, sig})$;
				\end{enumerate}
				\item Else
				\begin{enumerate}
					\setlength{\leftskip}{-6pt}
					\item \textbf{return};
				\end{enumerate}
			\end{enumerate}
			$\underline{\bf Reward}$ \\
			\vspace{1pt}
			\begin{itemize}
				\setlength{\leftskip}{-6pt}
				\item inputs: $\mathsf{tx_{open}}$, $\$_\mathsf{review}, \$_\mathsf{incentive}, \mathsf{acc_{pub}}, \mathsf{tsk}_i, k, \mathsf{sk}$
				\item outputs: $\mathsf{tx_{transfer}}$'s
			\end{itemize}
			\begin{enumerate}
				\setlength{\leftskip}{-6pt}
				\item For {each $\mathsf{reviewerID} \in \mathsf{tx_{open}.reviewerIDs}$}
				\begin{enumerate}
					\setlength{\leftskip}{-6pt}
					\item $\mathsf v = {\$_\mathsf{review}}$;
					\item $\mathsf{pk_r = GetPK(tx_{open}.reviewerID)}$;
					\item $\mathsf{tx_{origin}} = (\mathsf{acc_{pub}, pk_r,userID, v})$;
					\item $\mathsf{tsig}_i = \Pi_{ts}.\bf{ThresSign}$ $(\mathsf{tx_{origin}}, \mathsf{tsk}_i)$;
					\item send $\mathsf{tx_{origin}}$ to other validator to get other valid $\mathsf{tsig}_i$'s and store them in set $\mathsf{S_{tsig}}_i$;
					\item If {$\mathsf{GetNum(S_{{tsig}}}_i) \geq k$}
					\begin{enumerate}
						\setlength{\leftskip}{-6pt}
						\item $\mathsf{tsig} = {\Pi_{ts}.{\bf{SigShareComb}}(\mathsf{S_{tsig}}_i, k)}$;
						\item $\mathsf {tx_{transfer}} = (\mathsf{tx_{origin}, tsig})$;
						\item set the balance of $\mathsf{acc_{pub}}$ to $\mathsf{GetBalance(acc_{pub})}$ - $\mathsf{v}$;
						\item set the balance of $\mathsf{pk_r}$ to $\mathsf{GetBalance(pk_r)}$ + $\mathsf{v}$;
					\end{enumerate}
				\end{enumerate}
				\item If {$\mathsf{tx_{open}.result = accept}$}
				\begin{enumerate}
					\setlength{\leftskip}{-6pt}
					\item $\mathsf v = {\$_\mathsf{incentive}}$;
					\item $\mathsf{pk_r = GetPK(tx_{open}.userID)}$;
					\item $\mathsf{tx_{origin}} = (\mathsf{acc_{pub}, pk_r,userID, v})$;
					\item $\mathsf{tsig}_i = {\Pi_{ts}.\bf{ThresSign}(\mathsf {tx_{origin}, tsk}}_i)$;
					\item send $\mathsf{tx_{origin}}$ to other validator to get other valid $\mathsf{tsig}_i$'s and store them in set $\mathsf{S_{tsig}}_i$;
					\item If {$\mathsf{GetNum(S_{{tsig}}}_i) \geq k$}
					\begin{enumerate}
						\setlength{\leftskip}{-6pt}
						\item $\mathsf{tsig} = {\Pi_{ts}.{\bf{SigShareComb}}(\mathsf{S_{tsig}}_i, k)}$;
						\item $\mathsf {tx_{transfer}} = (\mathsf{tx_{origin}, tsig})$;
						\item set the balance of $\mathsf{acc_{pub}}$ to $\mathsf{GetBalance(acc_{pub})}$ - $\mathsf{v}$;
						\item set the balance of $\mathsf{pk_r}$ to $\mathsf{GetBalance(pk_r)}$ + $\mathsf{v}$;
					\end{enumerate}
				\end{enumerate}
				\item If {$\mathsf{GetDeposit(tx_{open}.userID) = 1}$/*$\mathsf{GetDeposit()}$ is a function to get the deposit state*/}
				\begin{enumerate}
					\setlength{\leftskip}{-6pt}
					\item $\mathsf v = {\$_\mathsf{deposit}}$;
					\item $\mathsf{pk_r = GetPK(tx_{open}.userID)}$;
					\item $\mathsf{tx_{origin}} = (\mathsf{acc_{pub}, pk_r,userID, v})$;
					\item $\mathsf{tsig}_i = {\Pi_{ts}.\bf{ThresSign}(\mathsf {tx_{origin}, tsk}}_i)$;
					\item send $\mathsf{tx_{origin}}$ to other validator to get other valid $\mathsf{tsig}_i$'s and store them in set $\mathsf{S_{tsig}}_i$;
					\item If {$\mathsf{GetNum(S_{{tsig}}}_i) \geq k$}
					\begin{enumerate}
						\setlength{\leftskip}{-6pt}
						\item $\mathsf{tsig} = {\Pi_{ts}.{\bf{SigShareComb}}(\mathsf{S_{tsig}}_i, k)}$;
						\item $\mathsf {tx_{transfer}} = (\mathsf{tx_{origin}, tsig})$;
						\item set the balance of $\mathsf{acc_{pub}}$ to $\mathsf{GetBalance(acc_{pub})}$ - $\mathsf{v}$;
						\item set the balance of $\mathsf{pk_r}$ to $\mathsf{GetBalance(pk_r)}$ + $\mathsf{v}$;
					\end{enumerate}
				\end{enumerate}
				\item \textbf{return} $\mathsf{tx_{transfer}}$'s;
			\end{enumerate}
			$\underline{\bf VerTx}$ \\
			\vspace{1pt}
			\begin{itemize}
				\setlength{\leftskip}{-6pt}
				\item inputs: $\mathsf{tx, mpk, grpID, acc_{pub}}$		
				\item outputs: {$b$}			
			\end{itemize}
			
			\begin{enumerate}
				\setlength{\leftskip}{-6pt}
				\item If {$\mathsf{tx = tx_{submit}}$}
				\begin{enumerate}
					\setlength{\leftskip}{-6pt}
					\setlength{\leftskip}{-6pt}
					\item $b = {\bf{TIBGS.Verify}(\mathsf{tx, tx.gsig, mpk, grpID})}$;
				\end{enumerate}
				\item ElseIf {$\mathsf{tx = tx_{distribute} || tx = tx_{review} || tx = tx_{open} }$}
				\begin{enumerate}
					\setlength{\leftskip}{-6pt}
					\item $b = {\Pi_{s}.\bf{Verify}(\mathsf{tx.pk_{sender}, tx, tx.sig})}$;
				\end{enumerate}
				\item ElseIf {$\mathsf{tx = tx_{transfer}}$}
				\begin{enumerate}
					\setlength{\leftskip}{-6pt}
					\item If {$\mathsf{\mathsf{GetBalance(tx.pk_{sender}) < tx.v}}$}
					\begin{enumerate}
						\setlength{\leftskip}{-6pt}
						\item \textbf{return} $0$;
					\end{enumerate}
					\item Else
					\begin{enumerate}
						\setlength{\leftskip}{-6pt}
						\item If {$\mathsf{tx.pk_{sender} = \mathsf{acc_{pub}}}$}
						\begin{enumerate}
							\setlength{\leftskip}{-6pt}
							\item $b = {\Pi_{ts}.\bf{Verify}(\mathsf{tx.pk_{sender}, tx, tx.tsig})}$;
						\end{enumerate}
						\item Else
						\begin{enumerate}
							\setlength{\leftskip}{-6pt}
							\item $b = {\Pi_{s}.\bf{Verify}(\mathsf{tx.pk_{sender}, tx, tx.sig})}$;
						\end{enumerate}
						\item If {$b=1$}
						\begin{enumerate}
							\setlength{\leftskip}{-6pt}
							\item set the balance of $\mathsf{tx.pk_{sender}}$ to $\mathsf{GetBalance}$ $\mathsf{(tx.pk_{sender})}$ - $\mathsf{tx.v}$;
							\item set the balance of $\mathsf{tx.pk_{receiver}}$ to $\mathsf{GetBalance}$ $\mathsf{(tx.pk_{receiver})}$ + $\mathsf{tx.v}$;
						\end{enumerate}
					\end{enumerate}
				\end{enumerate}
				\item \textbf{return} $b$;
			\end{enumerate}

		\end{multicols}
		\vspace{-0.55cm}
	\end{framed}
	
	\caption{The main algorithms of Open-Pub}\label{fig:alg}
	\vspace{-0.3cm}
\end{figure*}

\begin{enumerate}
	\setlength{\leftskip}{-10pt}
	\item {\bf System Initialization}($\lambda, k, n, \mathsf{grpID}$) $\rightarrow$ ($\mathsf{mpk}, \mathsf{msk}_i, \mathsf{gsk}_i,$ $\mathsf{gvk}_i, \mathsf{pk}, \mathsf{sk}, \mathsf{acc_{pub}}, \mathsf{tsk}_i, \mathsf{tvk}_i$).
	To initialize the system, validators that perform the duties of group managers run $\bf {TIBGS.Setup}$ and $\bf {TIBGS.GrpSetUp}$ with $\mathsf{grpID}$ to generate the master public key $\mathsf{mpk}$, the master private key share $\mathsf{msk}_i$, the group secret key share $\mathsf {gsk}_i$ and the group verify key share $\mathsf {gvk}_i$.
	Each validator initializes to create key pair $(\mathsf{pk,sk})$, and we use $\mathsf{pk}$ to represent the account.
	All validators run $\Pi_{ts}.\bf{Setup}$ and $\Pi_{ts}.\bf{ThresKeyGen}$ to generate a $(k,n)$ threshold signature account whose public key is $\mathsf{acc_{pub}}$ for storing deposits, review fees and incentive fees.
	The threshold signature secret key of each validator is $\mathsf {tsk}_i$, and the verification key is $\mathsf {tvk}_i$.
	Meanwhile, the system stipulates that the amount of deposit is $\$_\mathsf{deposit}$, the amount of review fee is $\$_\mathsf{review}$ and the amount of incentive fee is $\$_\mathsf{incentive}$.

	\item {\bf Registration}($\lambda, \mathsf{type}, \mathsf{userID}, \mathsf{acc_{pub}}, \mathsf{gsk}_i, \mathsf {gvk}_{i}, \$_\mathsf{deposit}, k$) $\rightarrow$ ($\mathsf{pk, sk}, \mathsf {tx_{transfer}}, \mathsf{usk}$).
	The system accepts the input of the $\mathsf{userID}$ and creates a key pair $(\mathsf{pk,sk})$ for various types of users. Users can sign up for three types of accounts: reader, reviewer, and author. A type identifier is used to distinguish these accounts, with a type $0$ for the reader, a type $1$ for the reviewer, and a type $2$ for the author. At the same time, authors need to pay $\$_\mathsf{deposit}$ to $\mathsf{acc_{pub}}$ through $\mathsf {tx_{transfer}}$ to prevent author from sabotaging the blockchain through anonymous transactions.
	A transfer transaction includes $(\mathsf{pk, acc_{pub}, userID, v, sig})$, where $\mathsf{v}$ is the amount of transfer and $\mathsf{sig}$ is the signature generated using $\mathsf{sk}$.
	After the deposit is confirmed, the validator will run $\bf{TIBGS.ExtShare}$ to generate $\mathsf{usk}_i$ and send it to the author, and the author can calculate the complete $\mathsf{usk}$ by $\bf{TIBGS.ReconstKey}$.

	\item {\bf Submission} ($\mathsf{field},$ $\mathsf{paper},$ $\mathsf{pk_{annoymity}},$ $ \mathsf{acc_{pub}},$ $\mathsf{usk}$) $\rightarrow$ $\mathsf{tx_{submit}}$.
	In the system, only the author account can submit the paper to the blockchain.
	An anonymous submit transaction includes $(\mathsf{pk_{annoymity}}, \mathsf{ acc_{pub}, field, paper, gsig})$.
	For anonymous transactions, the public key of the sender is $\mathsf{pk_{anonymity}}$.
	To generate an anonymous transaction, the author can run $\bf{TIBGS.Sign}$ with $\mathsf {usk}$ to generate a group signature $\mathsf {gsig}$.

	\item {\bf Distribution}($\mathsf{h_{tx_{submit}}},$ $\mathsf{reviewerIDs},$ $\mathsf{endtime},$ $ \mathsf{pk},$ $\mathsf{sk},$ $\mathsf{acc_{pub}}$) $\rightarrow$ $\mathsf{tx_{distribute}}$.
	The validator who successfully packages $\mathsf{tx_{submit}}$ into the block distributes the paper to the reviewers.
	When new $\mathsf{tx_{submit}}$ appears on the blockchain, the validator randomly selects reviewers based on the paper field.
	To hide the identity of reviewers, the validator encrypts $\mathsf{h_{tx_{submit}}}$, $\mathsf{reviewerID}$ and a random number $\mathsf{r}$ with the $\mathsf{pk}$ of the reviewer.
	The ciphertext is published as part of $\mathsf{tx_{distribute}}$, but only the corresponding reviewer can decrypt the ciphertext and get $\mathsf{h_{tx_{submit}}}$.
	In addition, this operation specifies the deadline for the review of the paper, which is denoted by $\mathsf{endtime}$.
	A distribute transaction includes $(\mathsf{pk, acc_{pub}, h_{tx_{submit}}, ciphertext..., endtime, sig})$, which the validator sends to $\mathsf{acc_{pub}}$.

	\item {\bf Review}($\mathsf{{reviewerID, tx_{distribute}, pk, sk, acc_{pub}}}$) $\rightarrow$ $\mathsf{tx_{review}}$.
	After $\mathsf{tx_{distribute}}$ is confirmed, reviewer can retrieve $\mathsf{tx_{distribute}}$ to find the corresponding ciphertext $\mathsf c$.
	By decrypting the ciphertext, reviewer can obtain plaintext including $\mathsf {{h_{tx_{submit}}}}$, $\mathsf{reviewerID}$ and $\mathsf{r}$.
	Through $\mathsf {{h_{tx_{submit}}}}$, the reviewer finds the paper in the database and reviews it.
	The reviewer will post comment and score through a review transaction including $(\mathsf{pk, acc_{pub}, h_{tx_{submit}}, reviewerID, r, comment, score, }\mathsf{sig})$.
	Until now, the author can know the true identity of this reviewer and the identities of reviewers who have not reviewed remain unknown.
	Readers can find papers and comments through $\mathsf{tx_{submit}}$ and $\mathsf{tx_{review}}$, and then review and score them through the review transaction.

	\item {\bf Open}($\mathsf{tx_{submit}},$ $\mathsf{tx_{distribute}},$ $ \mathsf{tx_{review}}$'s, $ \mathsf{result},$ $\mathsf{gsk}_i,$ $k$, $\mathsf{pk, sk}$,  $\mathsf{acc_{pub}}$) $\rightarrow$ $\mathsf{tx_{open}}$.
	After reaching the $\mathsf{endtime}$ of the paper, validator who distributes the paper will publish the author, the reviewers and the review result.
	Until then, the identity of the anonymous author has not been revealed.
	The distributor sends an open request and $\mathsf {tx_{submit}}$ to all validators, all of whom run $\bf {TIBGS.OpenPart}$ to generate $\mathsf{ok}_i$ and return it to the distributor.
	With at least $k$ $\mathsf{ok}_i$'s, the distributor runs $\bf {TIBGS.Open}$ to find the identity $\mathsf{userID}$ of the anonymous author.
	Finally, validator publishes the final result of the paper through the open transaction including $(\mathsf{pk, acc_{pub}, h_{tx_{submit}}, userID, result, reviewerIDs, sig})$.

	\item{\bf Reward}($\mathsf{tx_{open}},$ $\$_\mathsf{review},$ $\$_\mathsf{incentive},$ $ \mathsf{acc_{pub}},$ $\mathsf{tsk}_i,$ $k,$ $\mathsf{sk}$) $\rightarrow$ $\mathsf{tx_{transfer}}$'s.
	After the open operation, validators shall pay the review fee $\$_\mathsf{review}$ to reviewers according to the grade, and validators shall pay the incentive fee $\$_\mathsf{incentive}$ to author if the paper is accepted.
	These rewards will be paid out of account $\mathsf{acc_{pub}}$.
	If the deposit submitted by the author still exists, the deposit will be returned to the author.
	To decentralize power, a $\mathsf {tx_{transfer}}$ transaction transferred from $\mathsf{acc_{pub}}$ requires a threshold signature $\mathsf{tsig}$.
	A $\mathsf {tx_{transfer}}$ transaction from $\mathsf{acc_{pub}}$ includes $(\mathsf{acc_{pub}, pk_{receiver}, userID, v, tsig})$.
	The transaction passes verification only after at least $k$ validators have signed the transaction.

	\item{\bf VerTx}($\mathsf{tx, mpk, grpID, acc_{pub}}$) $\rightarrow$ $b$.
	Validators call this algorithm to check the validity of all types of transactions and then update the state of related accounts. The algorithm outputs $b = 1$ if tx is valid, otherwise it outputs $b = 0$.

\end{enumerate}

\section{Discussion and Analysis}
In this section, we first discuss the details of Open-Pub, then analyze its accountability, anonymity and recoverability.

\begin{table*}
	\vspace{-0.2cm}
	\center
	\caption{Comparison Between Open-Pub, Eureka, Orvium, PubChain and a Double-blind Review System}
	\vspace{-0.2cm}
	\label{table}
	\setlength{\tabcolsep}{3pt}
	\begin{tabular}{|p{100pt}|p{60pt}|p{60pt}|p{60pt}|p{60pt}|p{60pt}|p{60pt}|}
		\hline
		System        & Anonymity &  Traceability & Transparency & Validators/Editor & Author  & Reviewer   \\
		\hline
		Open-Pub      & Strong    & Yes  & Yes     &  Supervised   &Supervised & Supervised \\
		Eureka        & -    & Yes      & Yes     &  Supervised   &Supervised & Supervised \\
		Orvium        & -    & Yes           & Yes     &  Supervised   &Supervised & Supervised \\
		PubChain      & -    & Yes           & Yes     &  Supervised   &Supervised & Supervised \\
		Double-blind review system  & Weak    & No   & No   &  Unsupervised  &Unsupervised     & Unsupervised \\
		\hline
		\multicolumn{7}{p{500pt}}{Note: In Open-Pub, we do not have editors, and we authorize validators to distribute papers to reviewers, but validators have much less power than editors.}\\
	\end{tabular}
	\label{tab:compare}
	\vspace{-0.2cm}
\end{table*}

\subsection{Discussion}


{\bf Malicious participants.}
In Open-Pub, the participants of the single blockchain contain validators, authors, reviewers and readers, which cannot attack Open-Pub without loss of personal assets. There are reasons to explain the above occasions:
(i) the single blockchain in Open-Pub is a consortium blockchain with a secure consensus algorithm such as PBFT allowing that at most $f$ validators fail;
(ii) we require that each author should pay a deposit when registering an account, which is used to prevent authors from submitting malicious transactions;
(iii) if a reviewer forgers a review, he/she will lose his reputation;
(iv) readers’ comments do not affect the final results of the papers, and malicious comments will diminish their reputation.
Moreover, all operations of these malicious participants are recorded on the blockchain, and everyone can trace the related transactions to check the malicious operations.

{\bf Public account.}
A public account $\mathsf{acc_{pub}}$, which is maintained by all validators, is required to process deposits from authors and reward fees for reviewers and authors.
To reduce the risk of $\mathsf{acc_{pub}}$, we utilize the threshold signature to manage this account when any funds are transferred from this account. Before submitting a paper, the author should pay funds as a deposit to $\mathsf{acc_{pub}}$; after reviewing a valid paper, all validators control $\mathsf{acc_{pub}}$ to refund the deposit to the author and distribute rewards to the author and related reviewers. Note that we set the threshold for the threshold signature scheme to be the same as that of the TIBGS scheme to ensure its security.



{\bf Fair Review.} 
Using a decentralized system can prevent malicious behaviors from the centralized system. Due to the openness and transparency of the data, everyone can trace the whole process, which makes it difficult for validators and reviewers to act irrationally. We also design a strong double-blind mechanism to prevent validators and reviewers from being influenced by personal interests.
TIBGS hides the identity of the author from anyone when submitting multiple papers, which provides a stronger guarantee of fair review. Even in some cases where the identity of the author may be inferred from the text, public scrutiny can encourage reviewers to make as objective comments as possible. Since all comments will be published on the blockchain, obviously biased comments will be identified by the authors or other researchers easily.
In Open-Pub, public scrutiny and double-blind review work together to prevent misconducts and maintain the fairness of the review process.

{\bf Reward.}
In Open-Pub, validators reward reviewers and authors for their contributions, and these rewards are generated by the blockchain. In order to better motivate reviewers, it is important to distinguish between comments, such as more rewards for more serious comments. We evaluate the quality of comments from two aspects: paper citation after acceptance and other researchers’ opinions on these comments. The reward is based on the quality of the comment and the reviewer's past performance. The anonymity is the main goal of our current system, and we will investigate how to motivate authors and reviewers better using game theory as our future work. The evaluation mechanism and reward strategy will be designed as separate modules, which will be more easily integrated into Open-Pub.


\subsection{Analysis}
{\bf Accountability.}
In Open-Pub, only author accounts can generate anonymous transactions and the identity of the anonymous author will be disclosed later.
Anonymous transactions can cause the author to send spam transactions without being detected.
In addition to verifying the identity of the author, we require the author to pay a deposit upon registration.
Doing evil will cause the deposit to be locked up completely.

{\bf Anonymity.}
Open-Pub implements a double-blind review through TIBGS and asymmetric encryption.
Open-Pub implements anonymous transactions through TIBGS, and authors can hide their identities by publishing papers through anonymous transactions.
Only when the number of validators reaches the threshold can they collectively reveal the sender of the anonymous transaction.
In order to hide the identity of the reviewer, we do not send the transaction directly to the reviewer.
We will use the key of the reviewer to encrypt the identity of the reviewer and the paper information, and only the corresponding reviewer can decrypt the ciphertext.
The identity of the reviewer will be known to the author only after the reviewer has reviewed the paper.

{\bf Recoverability.}
The group private key $\mathsf{usk}$ is associated with the identity of the user, which has the advantage that the key can be recovered by reexecuting $\bf{TIBGS.ExtShare}$ and $\bf{TIBGS.ReconstKey}$ algorithms.
That is, validators whose quantity exceeds the threshold number can regenerate the group private key for the user in case of key loss.

\begin{mydef}[Full-anonymity]
	Let $\Theta = (\bf{SystemInitializ}\\
	\bf{ation}$, $\bf{Registration}$, $\bf{Submission}$, $\bf{Distribution}$, $\bf{Review}$,  $\bf{Open}$, $\bf{Reward}$, $\bf{VerTx})$ be the Open-Pub scheme.
	We say that $\Theta$ is fully anonymous if for all sufficiently large security parameter $k\in \mathbb{N}$ and any proper
	probabilistic polynomial time (PPT) adversary $\mathcal{A}$, its advantage
	$\mathsf{Adv}_{\Theta, \mathcal{A}}^{\mathsf{anon}}(1^\lambda)$ = $|\mathsf{Pr}[{\bf {Exp}}_{\Theta, \mathcal{A}}^{\mathsf{anon}}(1^\lambda)=1]$ - $\frac{1}{2}|$ is negligible.
\end{mydef}

\begin{mythm} Assuming that the TIBGS scheme is fully anonymous, the above Open-Pub scheme is also fully anonymous. (The proof is given in the Appendix.C.)
\end{mythm}

Based on the above analysis, we provide a comparison between Open-Pub, Eureka, Orvium, PubChain, and a traditional double-blind review system in Table~\ref{tab:compare}, in terms of anonymity, traceability, transparency, and participants. Open-Pub, Eureka, Orvium and PubChain all leverage blockchain technology for traceability and transparency, but only Open-Pub achieves strong anonymity on the blockchain. The existing double-blind review system has only weak anonymity, because the identities of anonymous authors and reviewers are known to the editor. In Open-Pub, the author's identity is protected by TIBGS, and even a single validator does not know the real identity, so it has strong anonymity.
\section{Implemention and Performance Evaluation}
In this section, we describe the implementation of Open-Pub, and then we present comprehensive experiment results to demonstrate its performance.
\subsection{Implemention}
We implement the Open-Pub system based on Ethereum source code in Golang language. Since Open-Pub is based on TIBGS, threshold signature and blockchain, our implementation mainly includes the following components:

{\em TIBGS}. To implement TIBGS, we use the PBC (Pairing-Based Cryptography) library which implements pairing-based cryptosystems in C language and a Go wrapper to use PBC. Based on PBC, we implement 8 algorithms of TIBGS including $\bf{Setup}$, $\bf{GrpSetUp}$, $\bf{ExtShare}$, $\bf{ReconstKey}$, $\bf{Sign}$, $\bf{Verify}$, $\bf{OpenPart}$ and $\bf{Open}$.

{\em Threshold signature}. We choose the threshold BLS \cite{boneh2001short} signature scheme as our threshold signature, and we implement it based on a Go library https://github.com/dfinity-side-projects/go-dfinity-crypto.

{\em PBFT and Threshold}. In Open-Pub, we adopt the PBFT algorithm as the consensus mechanism, and PBFT algorithm requires $3f + 1$ replicas to ensure security and activity in the case of $f$ failed nodes. We also need to set the threshold parameter of TIBGS and threshold signature, and we can set the threshold parameter as $(2f+1, 3f+1)$ to match the PBFT algorithm.

{\em Transactions}. In order to realize the function of paper review, we extend Ethereum by defining five types of transactions: $\mathsf{tx_{transfer}}$, $\mathsf{tx_{submit}}$, $\mathsf{tx_{distribute}}$, $\mathsf{tx_{review}}$, $\mathsf{tx_{open}}$. These transactions involve three signature algorithms including ECDSA signature \cite{johnson2001the}, TIBGS signature and threshold signature.
The $\mathsf{tx_{submit}}$ contains the TIBGS signature, the $\mathsf{tx_{transfer}}$ contains the threshold signature or ECDSA signature based on the different accounts and other types of transactions contains the ECDSA signature.

\subsection{Experiments and Performance}
Table~\ref{tab:key_size} shows the computation and communication costs of TIBGS. Except some constants, the computation and communication costs are determined by the threshold and the number of validators.

\begin{table}
	\vspace{-0.2cm}
	\center
	\caption{Computation and Communication Costs of TIBGS}
	\vspace{-0.2cm}
	\label{table}
	\setlength{\tabcolsep}{3pt}
	\begin{tabular}{|p{55pt}|p{100pt}|p{40pt}|}
		\hline
		Algorithm   &  Computation  &  Communication  \\
		\hline
		Setup & $2+k$ Exp & $O(n^2)$ \\
		GrpSetUp  & $6$ Exp & $O(n^2)$ \\
		ExtShare & $3$ Exp  & $-$ \\
		ReconstKey & $(4k+6)$ Exp + $9$ Pairing  & $O(k)$ \\
		Sign   & $1$ Pairing + $21$ Exp & $-$ \\
		Verify  & $3$ Pairing + $10$ Exp & $-$ \\
		OpenPart  & $2$ Pairing & $-$ \\
		Open  & $(2k+1)$ Exp & $O(k)$ \\
		\hline
		\multicolumn{3}{p{240pt}}{Note: Exp denotes exponentiation, Pairing denotes bilinear pairings, $k$ denotes threshold value and $n$ is the number of validators of Open-Pub.}\\
	\end{tabular}
	\label{tab:key_size}
	\vspace{-0.2cm}
\end{table}

\begin{figure*}[htbp]
	\vspace{-0.2cm}
	\centering
	\includegraphics[width=1\textwidth]{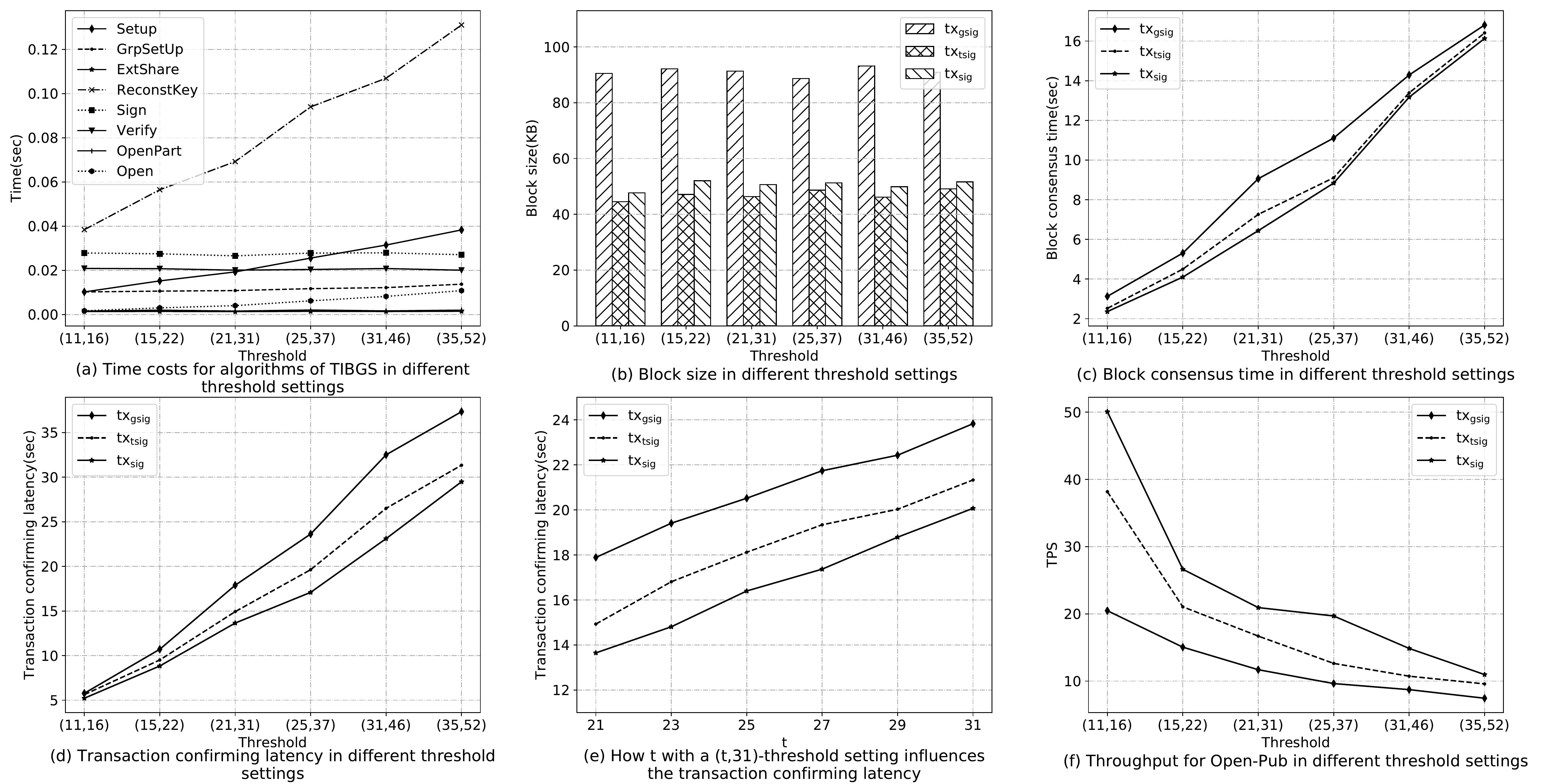}
	\vspace{-0.2cm}
	\caption{Experiment results for our implementation.}
	\label{fig:experiment}
	\vspace{-0.2cm}
\end{figure*}

To evaluate the performance of Open-Pub and the underlying TIBGS scheme, we deploy our system on $10$ Aliyun\footnote{https://www.aliyun.com/} ecs.g6.xlarge virtual machines, each of which has $4$ vCPU and $16$GB memory. We run $6$ docker containers on each virtual machine used to run the blockchain nodes independently. We measure the performance of TIBGS scheme and assess the impact of different thresholds on block size, block consensus time, transaction latency and throughput.



We first test the performance of the TIBGS scheme. We present the performance of each algorithm of TIBGS under different $(t, n)$-threshold.
We configure the threshold of Open-Pub to be $(11,16)$, $(15,22)$, $(21,31)$, $(25,37)$, $(31,46)$, $(35,52)$ respectively and set Open-Pub as $\mathsf{grpID}$.
As showed in Fig.~\ref{fig:experiment}. (a), the computation time for $\bf{Setup}$, $\bf{GrpSetUp}$, $\bf{ReconstKey}$ and $\bf{Open}$ increases steadily as the threshold increases, and $\bf{ExtShare, Sign, Verify}$ and $\bf{OpenPart}$ have almost fixed time costs. This is consistent with our analytical results summarized in Table~\ref{tab:key_size}.
It takes about $23$ ms for the author to sign anonymously and about $20$ ms for validators to verify the signature. Our TIBGS takes far less time than zero-knowledge proof algorithm zk-SNARKs, which needs tens of seconds to generate proofs.
TIBGS is a good choice to protect the identity of the author in the blockchain, and these costs are well worth the improvement in the fairness of the review process.

In the test, we divide all transactions into $\mathsf{tx_{sig}}$, $\mathsf{tx_{gsig}}$ and $\mathsf{tx_{tsig}}$ according to the type of signature, where $\mathsf{tx_{sig}}$ represents ECDSA signature transactions, $\mathsf{tx_{gsig}}$ represents TIBGS signature transactions, and $\mathsf{tx_{tsig}}$ represents threshold signature transactions. The signature size and verification time of the three signature algorithms are shown in Table~\ref{tab:size_time}.  We deploy some independent nodes to simulate the user node, which can generate $\mathsf{tx_{sig}}$, $\mathsf{tx_{gsig}}$ or $\mathsf{tx_{tsig}}$ in different thresholds settings.

\begin{table}[!ht]
	\vspace{-0.2cm}
	\center
	\caption{Signature Size and Verification Time}
	\vspace{-0.2cm}
	\label{table}
	\setlength{\tabcolsep}{3pt}
	\begin{tabular}{|p{95pt}|p{65pt}|p{65pt}|}
		\hline
		Signature   &  Signature size  &  Verification time  \\
		\hline
		ECDSA signature & $65$ bytes  & $0.2$ ms \\
		TIBGS Signature  & $533$ bytes & $20.1$ ms\\
		Threshold signature & $32$ bytes  & $0.7$ ms\\
		\hline
	\end{tabular}
	\label{tab:size_time}
	\vspace{-0.2cm}
\end{table}

At different thresholds, we measure the block size and the block consensus time when only one type of transaction is sent. Fig.~\ref{fig:experiment}. (b) shows that the block size does not change much under different thresholds.
The block size of all $\mathsf{tx_{gsig}}$ transactions is about $\mathsf{90KB}$, for all $\mathsf{tx_{tsig}}$ transactions it is about $\mathsf{48KB}$, and for all $\mathsf{tx_{sig}}$ transactions it is about $\mathsf{51KB}$.
The difference of block size under the same threshold mainly comes from the different size of ECDSA signature, threshold signature and TIBGS signature.
This indicates that the threshold does not affect the packaging process of the transaction.

Fig.~\ref{fig:experiment}. (c) shows that the block consensus time increases with the increase of the threshold. The block consensus time of $\mathsf{tx_{gsig}}$ transactions is the largest, and the block consistency time of $\mathsf{tx_{tsig}}$ and $\mathsf{tx_{sig}}$ is close.
As the threshold increases, the PBFT algorithm needs more time to consensus.
The difference of the block consensus time under the same threshold is related to the verification time of $\mathsf{tx_{gsig}}$, $\mathsf{tx_{tsig}}$ and $\mathsf{tx_{sig}}$.
In order to ensure the anonymity of $\mathsf{tx_{gsig}}$, the verification process of $\mathsf{tx_{gsig}}$ is much more complicated than that of $\mathsf{tx_{tsig}}$ and $\mathsf{tx_{sig}}$.

We measure the transaction confirming latency, which is the time between the transaction being issued by the user and being confirmed by Open-Pub.
Fig.~\ref{fig:experiment}. (d) shows that the transaction confirming latency of $\mathsf{tx_{gsig}}$ is greater than that of $\mathsf{tx_{tsig}}$ and $\mathsf{tx_{sig}}$, and both increase with the increase of threshold.
The increase of threshold will lead to the increase of consensus time, and naturally, the transaction confirming latency will increase.
The difference in transaction confirmation latency under the same threshold is due to the different validation times for the three types of signatures.
Under the condition of satisfying the PBFT algorithm, we set $t$ as $21, 23, 25, 27, 29, 31$ for a $(t, 31)$-threshold Open-Pub system. Fig.~\ref{fig:experiment}. (e) shows that the transaction confirmation latency is growing steadily as $t$ increases. The larger $t$ increases the block consensus time.

Fig.~\ref{fig:experiment}. (f) shows that the TPS (transactions per second) of the three types of transactions decreases as the threshold increases and the TPS of $\mathsf{tx_{sig}}$ is the maximum and TPS of $\mathsf{tx_{gsig}}$ is the minimum.
As the block consensus time increases with the threshold and the number of transactions per block remains roughly the same, TPS naturally declines.
The block consensus time of $\mathsf{tx_{gsig}}$ is the largest, so TPS of $\mathsf{tx_{gsig}}$ is the smallest.

Overall, the Open-Pub system has better performance for three types of transactions.
But the system has a slight performance degradation when handling $\mathsf{tx_{gsig}}$, which is the price of anonymity.
\section{Conclusion}

In this paper, we have presented Open-Pub, a transparent privacy-preserving academic publication system on blockchain.
In Open-Pub, we design a threshold group signature TIBGS, and we use TIBGS and asymmetric encryption to develop a strong double-blind mechanism to protect the identity of authors and reviewers.
In addition, we improve the transparency and fairness of the entire review process through blockchain.
We have analyzed the performance and security of Open-Pub and implemented Open-Pub based on Ethereum source code. Experimental results show that Open-Pub is highly efficient in dealing with anonymous transactions.

Future work can study appropriate incentive mechanisms to encourage the participation of authors, readers and reviewers.
Meanwhile, it may also be interesting to expand Open-Pub with more accurate metrics like impact factors for authors, reviewers, conferences and journals.

\section{Appendix}  

\subsection{Identity-based Group Signature}
An ID-based group signature IBGS scheme $\Lambda$ consists of six polynomial time algorithms
($\bf{Setup}, \bf{GrpSetUp}$, $\bf{Extract}, \bf{Sign}, \bf{Verify}, \bf{Open}$):
\begin{itemize}
	\item{${\bf{Setup}}(\mathsf1^{\lambda}) \rightarrow \mathsf{(mpk,msk)}$}. This algorithm generates a master public/private key pair $\mathsf{(mpk,msk)}$.
	
	\item{${\bf{GrpSetUp}}\mathsf{(grpID,msk)} \rightarrow \mathsf{gsk}$}. $\mathsf{GrpID}$ is a string that identifies the group. This algorithm on input of $\mathsf{grpID}$ and $\mathsf{msk}$ and outputs a group secret key $\mathsf{gsk}$. This $\mathsf{gsk}$ belongs to the group manager.
	\item{${\bf{Extract}}\mathsf{(userID,gsk)} \rightarrow \mathsf{usk}$}. The group manager executes the algorithm and outputs the group private key $\mathsf{usk}$, which is sent to the user.
	\item{${\bf{Sign}}(m,\mathsf{usk}) \rightarrow \mathsf{\sigma}$}. Each user can execute the algorithm and generate a signature $\mathsf\sigma$ corresponding to the message $m$.
	\item{${\bf{Verify}}\mathsf(m,\sigma,\mathsf{mpk,grpID}) \rightarrow \mathsf\{{0,1}\}$}. This algorithm can verify whether the signature is generated by user in the group.
	\item{${\bf{Open}}\mathsf(\mathsf{gsk},\sigma,m) \rightarrow \mathsf{userID}$}. The group manager can execute the algorithm and reveal the identifier $\mathsf{userID}$ of the user who produced the signature $\mathsf\sigma$ corresponding to the message $m$.
\end{itemize}

\noindent {\bf Security Model.} We recall the security model defined by Smart and Warinschi \cite{smart2009identity} for the identity-based group signature case.
The security model defines two security notions, namely {\em full-anonymity} and {\em full-traceability}.
{\em Full-anonymity} captures the anonymity property of the TIBGS scheme by an indistinguishability
experiment between an adversary and the group signature scheme, while {\em full-traceability} captures
the traceability property by a traceability experiment between an adversary and the group signature scheme.

The full-anonymity experiment for the IBGS scheme defined in \cite{smart2009identity} is defined in Fig.~\ref{fig:anon1}:
\begin{figure}[!ht]
	\begin{framed} \small
		\vspace{-0.2cm}
		\raggedright
		
		$\underline{{\bf {Exp}}_{\Lambda, \mathcal{A}}^{\mathsf{anon}}(1^\lambda):}$\\
		\setlength{\leftskip}{10pt}$(\mathsf{mpk,msk})\leftarrow {\bf {Setup}}(1^\lambda)$\\
		$(\mathrm{grpID}^*, \mathrm{userID}_0, \mathrm{userID}_1, m, \mathsf{state})\leftarrow \mathcal{A}_1^{\mathbf{GrpSetup(\cdot), Extract(\cdot),Open(\cdot)}}(\mathsf{mpk})$\\
		$b \stackrel{\$}{\leftarrow} \{0,1\}$\\
		$\sigma^*\leftarrow \mathsf{Sign}(m,\mathsf{usk}),$ where $((\mathrm{grpID}^*,\mathrm{userID}_b),\mathsf{usk})$ $\in$ ${\bf {userID}s}$\\
		$b'\leftarrow \mathcal{A}_2^{\mathbf{GrpSetup(\cdot), Extract(\cdot),Open(\cdot)}}(\sigma^*,\mathsf{state})$\\
		if $b'=b$ return 1\\
		else return 0\\
		
		\vspace{0.3cm}
		\setlength{\leftskip}{0pt}$\underline{\mathcal{O}^{\bf {GrpSetUp}}_{\mathsf{msk}}(\mathrm{grpID})}:$\\
		\setlength{\leftskip}{10pt}if $\exists$ $(\mathrm{grpID}, \mathsf{gsk})$ $\in$ ${\bf {grpID}s}$\\
		\setlength{\leftskip}{20pt}return $\mathsf{gsk}$\\
		\setlength{\leftskip}{10pt}else $\mathsf{gsk}\leftarrow {\bf {GrpSetUp}}(\mathsf{msk}, \mathrm{grpID})$\\
		\setlength{\leftskip}{20pt}return $\mathsf{gsk}$\\
		
		\vspace{0.3cm}
		\setlength{\leftskip}{0pt}$\underline{\mathcal{O}^{\bf {Extract}}_{\mathsf{msk,gsk}}(\mathrm{grpID},\mathrm{userID})}:$\\
		\setlength{\leftskip}{10pt}if $\nexists$ $(\mathrm{grpID}, \mathsf{gsk})$ $\in$ ${\bf {grpID}s}$\\
		\setlength{\leftskip}{20pt}$\mathsf{gsk}\leftarrow {\bf {GrpSetUp}}(\mathsf{msk}, \mathrm{grpID})$\\
		\setlength{\leftskip}{10pt}if $\nexists$ $((\mathrm{grpID},\mathrm{userID}), \mathsf{usk})$ $\in$ ${\bf {userID}s}$\\
		\setlength{\leftskip}{20pt}$\mathsf{usk}\leftarrow {\bf {Extract}}(\mathsf{gsk}, \mathrm{userID})$\\
		\setlength{\leftskip}{10pt}return $\mathsf{usk}$\\
		
		\vspace{0.3cm}
		\setlength{\leftskip}{0pt}$\underline{\mathcal{O}^{\bf {Open}}_{\mathsf{gsk}}(\mathrm{grpID},\sigma, m)}:$\\
		\setlength{\leftskip}{10pt}if $\exists$ $(\mathrm{grpID}, \mathsf{gsk})$ $\in$ ${\bf {grpIDs}}$\\
		\setlength{\leftskip}{20pt}return $\mathrm{userID}\leftarrow {\bf {Open}}(\mathsf{gsk}, \sigma, m)$\\
		\setlength{\leftskip}{10pt}else return $\perp$\\
		
	\end{framed}
	
	\caption{The full-anonymity experiment for IBGS in \cite{smart2009identity}.}\label{fig:anon1}
	\vspace{-0.3cm}
\end{figure} \normalsize

\begin{mydef}[Full-anonymity]
	Let $\Lambda = (\bf{Setup}$, $\bf{GrpSetUp}$, $\bf{Extract}$, $\bf{Sign}$, $\bf{Verify}$, $\bf{Open})$ be an identity-based group signature scheme.
	We say that $\Lambda$ is fully-anonymous if for all sufficiently large security parameter $k\in \mathbb{N}$ and any proper
	probabilistic polynomial time (PPT) adversary $\mathcal{A}$, its advantage
	$\mathsf{Adv}_{\Lambda, \mathcal{A}}^{\mathsf{anon}}(1^\lambda)$ = $|\mathsf{Pr}[{\bf Exp}_{\Lambda, \mathcal{A}}^{\mathsf{anon}}(1^\lambda)=1]$ - $\frac{1}{2}|$ is negligible.
\end{mydef}

It has been proved in \cite{smart2009identity} that the IBGS in \cite{smart2009identity} is fully anonymous.
We omit the full-traceability experiment and the corresponding
theorem for conciseness.

\subsection{Threshold Identity-based Group Signature}

The full-anonymity experiment is defined in Fig.~\ref{fig:anon}. The adversary is allowed to query several oracles,
$\mathbf{GrpSetUp}$, $\mathbf{ExtShare}$ and $\mathbf{OpenPart}$. The adversary generates a group identity and two user identities for which it will be challenged with a signature signed by one of the users. TIBGS achieves full anonymity if the adversary fails to guess the correct user identity with non-negligible probability.

\begin{figure}[!ht]
	\begin{framed} \small
		\vspace{-0.2cm}
		\raggedright
		
		$\underline{{\bf {Exp}}_{\Pi, \mathcal{A}}^{\mathsf{anon}}(1^\lambda):}$\\
		\setlength{\leftskip}{10pt}$(\mathsf{mpk,msk_i})\leftarrow {\bf {Setup}}(1^\lambda)$\\
		$(\mathsf{grpID}^*, \mathsf{userID}_0, \mathsf{userID}_1, m, \mathsf{state})\leftarrow$\\
		\setlength{\leftskip}{20pt}$\mathcal{A}_1^{\mathbf{GrpSetup, ExtShare,OpenPart}}(\mathsf{mpk})$\\
		\setlength{\leftskip}{10pt}$b \stackrel{\$}{\leftarrow} \{0,1\}$\\
		$\sigma^*\leftarrow \bf{Sign}(m,\mathsf{usk}),$ where $\mathsf{usk}\leftarrow \{\mathsf{usk}_i\}_{i\in \mathbf{S}}$ \\
		\setlength{\leftskip}{20pt}and $((\mathsf{grpID}^*, i, \mathsf{userID}_b),\mathsf{usk}_i)$ $\in$ ${\bf {userID}s}$  \\
		\setlength{\leftskip}{10pt}$b'\leftarrow \mathcal{A}_2^{\mathbf{GrpSetup, ExtShare,OpenPart}}(\sigma^*,\mathsf{state})$\\
		if $b'=b$ return 1\\
		else return 0\\
		
		\vspace{0.3cm}
		\setlength{\leftskip}{0pt}$\underline{\mathcal{O}^{\bf {GrpSetUp}}_{\mathsf{msk_i}}(\mathsf{grpID},i)}:$\\
		\setlength{\leftskip}{10pt}if $\exists$ $(\mathsf{grpID}, i, \mathsf{gsk}_i, \mathsf{gvk}_i)$ $\in$ ${\bf {grpID}s}$\\
		\setlength{\leftskip}{20pt}return $\mathsf{gsk}_i, \mathsf{gvk}_i$\\
		\setlength{\leftskip}{10pt}else $\mathsf{gsk}_i, \mathsf{gvk}_i\leftarrow {\bf {GrpSetUp}}(\mathsf{grpID}, i,  \mathsf{msk_i})$\\
		\setlength{\leftskip}{20pt}return $\mathsf{gsk}_i, \mathsf{gvk}_i$\\
		
		\vspace{0.3cm}
		\setlength{\leftskip}{0pt}$\underline{\mathcal{O}^{\bf {ExtShare}}_{\mathsf{msk,gsk}_i}(\mathsf{grpID}, i, \mathsf{userID})}:$\\
		\setlength{\leftskip}{10pt}if $\nexists$ $(\mathsf{grpID}, i, \mathsf{gsk}_i)$ $\in$ ${\bf {grpID}s}$\\
		\setlength{\leftskip}{20pt}$\mathsf{gsk}_i\leftarrow {\bf {GrpSetUp}}(\mathsf{grpID}, i, \mathsf{msk})$\\
		\setlength{\leftskip}{10pt}if $\nexists$ $((\mathsf{grpID}, i, \mathsf{userID}), \mathsf{usk}_i)$ $\in$ ${\bf {userID}s}$\\
		\setlength{\leftskip}{20pt}$\mathsf{usk}_i\leftarrow {\bf {ExtShare}}(\mathsf{gsk}_i, \mathsf{userID})$\\
		\setlength{\leftskip}{10pt}return $\mathsf{usk}_i$\\
		
		\vspace{0.3cm}
		\setlength{\leftskip}{0pt}$\underline{\mathcal{O}^{\bf {OpenPart}}_{\mathsf{gsk}_i}(\mathsf{grpID}, i, \sigma, m)}:$\\
		\setlength{\leftskip}{10pt}if $\exists$ $(\mathsf{grpID}, i, \mathsf{gsk}_i)$ $\in$ ${\bf {grpID}s}$\\
		\setlength{\leftskip}{20pt}return $\mathsf{ok}_i\leftarrow {\bf {OpenPart}}(\mathsf{gsk}_i, \sigma, m)$\\
		\setlength{\leftskip}{10pt}else return $\perp$\\
		
	\end{framed}
	
	\caption{The full anonymity experiment for TIBGS. It maintains two lists: $\mathbf{grpIDs}$ contains all group identities with their private keys, and $\mathbf{userIDs}$ contains all user identities with their private keys.
		$\mathbf{S}_{\mathsf{grpID}^*}$ represents the index set of the group managers.}\label{fig:anon}
	\vspace{-0.3cm}
\end{figure} \normalsize

The full-traceability experiment is defined in Fig.~\ref{fig:trace}. Similar to the full-anonymity experiment, the adversary
is also allowed to query several oracles,
$\mathbf{GrpSetUp}$, $\mathbf{ExtShare}$, $\mathbf{Sign}$ and $\mathbf{OpenPart}$. The adversary generates a group identity and
a signature of a message $m$. TIBGS achieves full traceability
if the signature produced by the adversary cannot be traced to one of the corrupted users with negligible probability.
\begin{figure}[!ht]
	\begin{framed} \small
		\vspace{-0.2cm}
		\raggedright
		$\underline{{\bf {Exp}}_{\Pi, \mathcal{A}}^{\mathsf{trace}}(1^\lambda):}$\\
		\setlength{\leftskip}{10pt}$(\mathsf{mpk,msk})\leftarrow {\bf {Setup}}(1^\lambda)$\\
		$(m, \sigma, \mathsf{grpID^*})\leftarrow \mathcal{A}_1^{\mathbf{GrpSetup, ExtShare,OpenPart}}(\mathsf{mpk})$\\
		let $\mathsf{gsk}^*_i = \mathbf{GrpSetUp}(\mathsf{grpID}^*, i, \mathsf{msk})$ for $i \in \mathbf{S}_{\mathsf{grpID}^*}$\\
		$\mathsf{ok}_i\leftarrow \mathbf{OpenPart}(\mathsf{gsk}_i^*, \sigma, m)$\\
		$\mathsf{userID}\leftarrow \mathbf{Open}(\{\mathsf{ok}_i\}_{i\in \mathbf{S}_{\mathsf{grpID}^*}})$\\
		if $\mathbf{Verify}(m, \sigma, \mathsf{mpk}, \mathsf{grpID^*})$= $\mathtt{false}$ or $(\mathsf{grpID^*},i, \mathsf{userID})$ \\
		\setlength{\leftskip}{20pt}$\in$ $\mathbf{corrgrpIDs}$ for at most $t-1$ different $i$\\
		\setlength{\leftskip}{20pt}return 0 \\
		\setlength{\leftskip}{10pt}else return 1
		
		\vspace{0.3cm}
		\setlength{\leftskip}{0pt}$\underline{\mathcal{O}^{\bf {GrpSetUp}}_{\mathsf{msk}}(\mathsf{grpID},i)}:$\\
		\setlength{\leftskip}{10pt}if $\exists$ $(\mathsf{grpID}, i, \mathsf{gsk}_i, \mathsf{gvk}_i)$ $\in$ ${\bf {grpID}s}$\\
		\setlength{\leftskip}{20pt}return $\mathsf{gsk}_i, \mathsf{gvk}_i$\\
		\setlength{\leftskip}{10pt}else $\mathsf{gsk}_i, \mathsf{gvk}_i\leftarrow {\bf {GrpSetUp}}(\mathsf{grpID}, i,  \mathsf{msk})$\\
		\setlength{\leftskip}{20pt}return $\mathsf{gsk}_i, \mathsf{gvk}_i$\\
		
		\vspace{0.3cm}
		\setlength{\leftskip}{0pt}$\underline{\mathcal{O}^{\bf {ExtShare}}_{\mathsf{msk,gsk}_i}(\mathsf{grpID}, i, \mathsf{userID}, \mathsf{type})}:$\\
		\setlength{\leftskip}{10pt}if $\nexists$ $(\mathsf{grpID}, i, \mathsf{gsk}_i)$ $\in$ ${\bf {grpID}s}$\\
		\setlength{\leftskip}{20pt}$\mathsf{gsk}_i\leftarrow {\bf {GrpSetUp}}(\mathsf{grpID}, i, \mathsf{msk})$\\
		\setlength{\leftskip}{10pt}if $\nexists$ $((\mathsf{grpID},i,\mathsf{userID}), \mathsf{usk}_i)$ $\in$ ${\bf {userID}s}$\\
		\setlength{\leftskip}{20pt}$\mathsf{usk}_i\leftarrow {\bf {Extract}}(\mathsf{gsk}_i, \mathsf{userID})$\\
		\setlength{\leftskip}{10pt}if $\mathsf{type}=\mathtt{corrupt}$\\
		\setlength{\leftskip}{20pt}add $(\mathsf{grpID}, i, \mathsf{userID})$ to $\mathbf{corrgrpIDs}$\\
		\setlength{\leftskip}{10pt}return $\mathsf{usk}_i$\\
		
		\vspace{0.3cm}
		\setlength{\leftskip}{0pt}$\underline{\mathcal{O}^{\bf {Sign}}_{\mathsf{usk}}(\mathsf{grpID}, \mathsf{userID}, m)}:$\\
		\setlength{\leftskip}{10pt}if $\exists ((\mathsf{grpID}, i, \mathsf{userID}),\mathsf{usk}_i)$ $\in$ ${\bf {userID}s}$\\
		\setlength{\leftskip}{20pt}$\mathsf{usk}\leftarrow \mathbf{ReconstKey}(\mathsf{userID},\{\mathsf{usk}_i\}_{i\in \mathbf{S}})$\\
		\setlength{\leftskip}{20pt}$\sigma\leftarrow \mathbf{Sign}(m, \mathsf{usk})$\\
		\setlength{\leftskip}{20pt}return $\sigma$\\
		\setlength{\leftskip}{10pt}else return $\perp$
		
		\vspace{0.3cm}
		\setlength{\leftskip}{0pt}$\underline{\mathcal{O}^{\bf {OpenPart}}_{\mathsf{gsk}_i}(\mathsf{grpID}, i, \sigma, m)}:$\\
		\setlength{\leftskip}{10pt}if $\exists$ $(\mathsf{grpID}, i, \mathsf{gsk}_i)$ $\in$ ${\bf {grpID}s}$\\
		\setlength{\leftskip}{20pt}return $\mathsf{ok}_i\leftarrow {\bf {OpenPart}}(\mathsf{gsk}_i, \sigma, m)$\\
		\setlength{\leftskip}{10pt}else return $\perp$\\
		
	\end{framed}
	
	\caption{The full traceability experiment for TIBGS. It maintains three lists: $\mathbf{corrgrpIDs}$ contains the corrupted user identities, $\mathbf{grpIDs}$ contains all group identities with their private keys, and $\mathbf{userIDs}$ contains all user identities with their private keys.
		$\mathbf{S}_{\mathsf{grpID}^*}$ represents the index set of the group managers.}\label{fig:trace}
	\vspace{-0.3cm}
\end{figure} \normalsize

\begin{proof}
	We reduce the full anonymity of our TIBGS scheme (denoted as $\Pi$) to that of the IBGS scheme (denoted as $\Lambda$) in~\cite{smart2009identity}. Suppose there is
	a polynomial-time adversary $\mathcal{A}$ can break the full anonymity of $\Pi$, we construct another adversary $\mathcal{B}$ that uses
	$\mathcal{A}$ as a subroutine to break the full anonymity of $\Lambda$.
	
	The challenger $\mathcal{C}$ of $\Lambda$ executes $\mathbf{Setup}$ to output $\mathsf{mpk}$ and gives it to
	$\mathcal{B}$ as in Fig.~\ref{fig:anon1}. Then $\mathcal{B}$ passes $\mathsf{msk}$ to
	$\mathcal{A}$. As per the full-anonymity experiment, $\mathcal{A}$ makes the following oracle queries, which are answered
	by $\mathcal{B}$ as follows:
	\begin{itemize}
		\item $\mathcal{O}_{\mathsf{msk}}^{\mathbf{GrpSetUp}}(\mathsf{grpID},i)$: If $\mathsf{grpID}\neq \mathsf{grpID^*}$,
		$\mathcal{B}$ obtains $\mathsf{gsk}$ by querying the oracle $\mathcal{O}_{\mathsf{msk}}^{\mathbf{GrpSetUp}}(\mathsf{grpID})$
		in Fig.~\ref{fig:anon1}. Then $\mathcal{B}$ computes $\mathsf{gsk}_i$ as well as $\mathsf{gvk}_i$ from $\mathsf{gsk}$ and $i$ for $\mathcal{A}$. Otherwise, $\mathcal{B}$ randomly generates $\mathsf{gsk}_i$ and $\mathsf{gvk}_i$ for $\mathcal{A}$.
		
		\item $\mathcal{O}_{\mathsf{msk,gsk}_i}^{\mathbf{ExtShare}}(\mathsf{grpID},i,\mathsf{userID})$: If $\mathsf{userID}\neq \mathsf{userID}_0$ or $\mathsf{userID}_1$, $\mathcal{B}$ obtains $\mathsf{usk}$ by querying the oracle $\mathcal{O}_{\mathsf{msk,gsk}}^{\mathbf{ExtShare}}(\mathsf{grpID},\mathsf{userID})$
		in Fig.~\ref{fig:anon1}. Then $\mathcal{B}$ computes $\mathsf{usk}_i$ from $\mathsf{usk}$ and $i$ for $\mathcal{A}$.
		Otherwise, $\mathcal{B}$ randomly generates $\mathsf{usk}_i$ for $\mathcal{A}$.
		
		\item $\mathcal{O}_{\mathsf{gsk}_i}^{\mathbf{OpenPart}}(\mathsf{grpID},i,\sigma,m)$: If $\mathsf{grpID}\neq \mathsf{grpID^*}$, $\mathcal{B}$ obtains $\mathsf{gsk}$ by querying the oracle
		$\mathcal{O}_{\mathsf{msk}}^{\mathbf{GrpSetUp}}(\mathsf{grpID})$. Then $\mathcal{B}$ computes $\mathsf{gsk}_i$ as well as $\mathsf{gvk}_i$ from $\mathsf{gsk}$ and $i$. After that, $\mathcal{B}$ executes $\mathbf{OpenPart}(\mathsf{gsk}_i, \sigma, m)$ and return the result to $\mathcal{A}$. Otherwise, $\mathcal{B}$ randomly generates $\mathsf{ok}_i$ for $\mathcal{A}$.
	\end{itemize}
	
	After $\mathcal{A}$ has made enough oracle queries, $\mathcal{A}$ outputs $(\mathsf{grpID}^*, \mathsf{userID}_0, \mathsf{userID}_1, m, \mathsf{state})$ to $\mathcal{B}$, who will forward the output to the challenger $\mathcal{C}$. Then $\mathcal{C}$
	outputs a signature $\sigma^*$ to $\mathcal{B}$ who forwards $\sigma^*$ to $\mathcal{A}$ and obtains the output $b'$
	from $\mathcal{A}$. Finally, $\mathcal{B}$ outputs $b'$ as its guess for $b$ chosen by $\mathcal{C}$.
	
	Clearly, the adversary $\mathcal{B}$ has the same advantage of the experiment as $\mathcal{A}$, i.e.,
	\[\mathsf{Adv}_{\Lambda, \mathcal{B}}^{\mathsf{anon}}(1^\lambda) = \mathsf{Adv}_{\Pi, \mathcal{A}}^{\mathsf{anon}}(1^\lambda).\]
	Since no
	such adversary $\mathcal{B}$ can break full anonymity of $\Lambda$, we conclude that $\mathcal{A}$ cannot break full anonymity of $\Pi$.
	
\end{proof}


\subsection{Open-Pub}

The full-anonymity experiment for Open-Pub is defined in Fig.~\ref{fig:anon3}.
The adversary is allowed to query several oracles including $\mathbf{AuthorRegistration}$ and $\mathbf{Open}$.
The adversary generates two user identities for which it will be challenged with a signature signed by one of the users. Open-Pub achieves full anonymity if the adversary fails to guess the correct user identity with non-negligible probability.
\begin{figure}[!ht]
	\begin{framed} \small
		\vspace{-0.2cm}
		\raggedright
		
		$\underline{{\bf {Exp}}_{\Theta, \mathcal{A}}^{\mathsf{anon}}(1^\lambda):}$\\
		\setlength{\leftskip}{10pt}$(\mathsf{mpk},\mathsf{msk}_i,\mathsf{gsk}_i,\mathsf{gvk}_i)\leftarrow$\\
		\setlength{\leftskip}{20pt}${\bf {SystemInitialization}}(1^\lambda, \mathsf{grpID})$\\
		\setlength{\leftskip}{10pt}$(\mathrm{userID}_0, \mathrm{userID}_1, \mathrm{h_{paper}}, \mathsf{state})\leftarrow$\\ \setlength{\leftskip}{20pt}$\mathcal{A}_1^{\mathbf{AuthorRegistration(\cdot), Open(\cdot)}}(\mathsf{mpk},\mathsf{msk}_i,\mathsf{gsk}_i,\mathsf{gvk}_i)$\\
		\setlength{\leftskip}{10pt}$b \stackrel{\$}{\leftarrow} \{0,1\}$\\
		$\mathsf{gsig}^*\leftarrow \bf{Submit}(\mathrm{h_{paper}},\mathsf{usk}),$ where $((\mathrm{userID}_b),\mathsf{usk})$ $\in$ ${\bf{userID}s}$\\
		$b'\leftarrow \mathcal{A}_2^{\mathbf{AuthorRegistration(\cdot),Open(\cdot)}}(\mathsf{gsig}^*,\mathsf{state})$\\
		if $b'=b$ return 1\\
		else return 0\\

		\vspace{0.3cm}
		\setlength{\leftskip}{0pt}$\underline{\mathcal{O}^{\bf {AuthorRegistration}}_{\mathsf{msk}_i,\mathsf{gsk}_i}(\mathrm{grpID}, i, \mathrm{userID})}:$\\
		\setlength{\leftskip}{10pt}if $\exists$ $(\mathrm{grpID}, \mathsf{gsk}_i)$ $\in$ ${\bf {grpIDs}}$\\
		\setlength{\leftskip}{20pt}$\mathsf{gsk}_i, \mathsf{gvk}_i\leftarrow {\bf {GrpSetUp}}(\mathsf{grpID}, i, \mathrm{msk}_i)$\\
		\setlength{\leftskip}{10pt}if $\nexists$ $((\mathrm{userID}), \mathsf{usk})$ $\in$ ${\bf {userID}s}$\\
		\setlength{\leftskip}{20pt}$\mathsf{usk}_i\leftarrow {\bf {ExtShare}}(\mathsf{gsk}_i, \mathrm{userID})$\\
		\setlength{\leftskip}{20pt}$\mathsf{usk}\leftarrow {\bf {ReconstKey}}(\mathsf{usk}_i, \mathrm{userID})$\\
		\setlength{\leftskip}{10pt}return $\mathsf{usk}$\\
		
		\vspace{0.3cm}
		\setlength{\leftskip}{0pt}$\underline{\mathcal{O}^{\bf {Open}}_{\mathsf{gsk}_i}(\mathrm{grpID}, i, \mathsf{gsig}, \mathsf{h_{paper}})}:$\\
		\setlength{\leftskip}{10pt}if $\exists$ $(\mathrm{grpID}, \mathsf{gsk}_i)$ $\in$ ${\bf {grpIDs}}$\\
		\setlength{\leftskip}{20pt}$\mathrm{ok}_i\leftarrow {\bf{OpenPart}}(\mathsf{gsk}_i, \mathsf{gsig}, \mathsf{h_{paper}})$\\
		\setlength{\leftskip}{20pt}return $\mathrm{userID}\leftarrow {\bf{Open}}(\mathsf{ok}_i)$\\
		\setlength{\leftskip}{10pt}else return $\perp$\\
		
	\end{framed}
	
	\caption{The full-anonymity experiment for Open-Pub.}\label{fig:anon3}
	\vspace{-0.3cm}
\end{figure} \normalsize

\begin{proof}
	We reduce the full anonymity of the Open-Pub scheme (denoted as $\Theta$) to that of the TIBGS scheme (denoted as $\Pi$). Suppose there is
	a polynomial-time adversary $\mathcal{A}$ can break the full anonymity of $\Theta$, we construct another adversary $\mathcal{B}$ that uses
	$\mathcal{A}$ as a subroutine to break the full anonymity of $\Pi$.
	
	The challenger $\mathcal{C}$ of $\Theta$ executes $\mathbf{SystemInitialization}$ to output a tuple $(\mathsf{mpk},\mathsf{msk}_i, \mathsf{gsk}_i, \mathsf{gvk}_i)$ and gives it to
	$\mathcal{B}$ as in Fig.~\ref{fig:anon3}. Then $\mathcal{B}$ passes $(\mathsf{mpk},\mathsf{msk}_i, \mathsf{gsk}_i, \mathsf{gvk}_i)$ to
	$\mathcal{A}$. As per the full-anonymity experiment, $\mathcal{A}$ makes the following oracle queries, which are answered
	by $\mathcal{B}$ as follows:
	\begin{itemize}
		
		\item $\mathcal{O}_{\mathsf{msk}_i, \mathsf{gsk}_i}^{\mathbf{Author Registration}}(\mathsf{grpID},i,\mathsf{userID})$: If $\mathsf{userID}\neq \mathsf{userID}_0$ or $\mathsf{userID}_1$, $\mathcal{B}$ obtains $\mathsf{usk}_i$ by querying the oracle $\mathcal{O}_{\mathsf{msk}_i, \mathsf{gsk}_i}^{\mathbf{ExtShare}}(\mathsf{grpID}, i, \mathsf{userID})$
		in Fig.~\ref{fig:anon}. Then $\mathcal{B}$ computes $\mathsf{usk}$ from $\mathsf{usk}_i$ and $i$ for $\mathcal{A}$.
		Otherwise, $\mathcal{B}$ randomly generates $\mathsf{usk}$ for $\mathcal{A}$.
		
		\item $\mathcal{O}_{\mathsf{gsk}_i}^{\mathbf{Open}}(\mathsf{grpID},i,\mathsf{gsig},\mathsf{h_{paper}})$: If $\mathsf{grpID}\neq \mathsf{grpID^*}$, $\mathcal{B}$ obtains $(\mathsf{gsk}_i, \mathsf{gvk}_i)$ by querying the oracle
		$\mathcal{O}_{\mathsf{msk}}^{\mathbf{GrpSetUp}}(\mathsf{grpID}, i)$. After that, $\mathcal{B}$ executes $\mathbf{OpenPa}$ $\mathbf{rt}(\mathsf{gsk}_i, \mathsf{gsig},\mathsf{h_{paper}})$ to get $\mathsf{ok}_i$, and then executes $\mathbf{Open}(\mathsf{ok}_i )$ and return the result to $\mathcal{A}$. Otherwise, $\mathcal{B}$ randomly generates $\mathsf{ok}$ for $\mathcal{A}$.
	\end{itemize}
	
	After $\mathcal{A}$ has made enough oracle queries, $\mathcal{A}$ outputs $(\mathsf{grpID}^*, \mathsf{userID}_0, \mathsf{userID}_1, \mathsf{h_{paper}}, \mathsf{state})$ to $\mathcal{B}$, who will forward the output to the challenger $\mathcal{C}$. Then $\mathcal{C}$
	outputs a signature $\mathsf{gsig}^*$ to $\mathcal{B}$ who forwards $\mathsf{gsig}^*$ to $\mathcal{A}$ and obtains the output $b'$
	from $\mathcal{A}$. Finally, $\mathcal{B}$ outputs $b'$ as its guess for $b$ chosen by $\mathcal{C}$.
	
	Clearly, the adversary $\mathcal{B}$ has the same advantage of the experiment as $\mathcal{A}$, i.e.,
	\[\mathsf{Adv}_{\Pi, \mathcal{B}}^{\mathsf{anon}}(1^\lambda) = \mathsf{Adv}_{\Theta, \mathcal{A}}^{\mathsf{anon}}(1^\lambda).\]
	Since no
	such adversary $\mathcal{B}$ can break full anonymity of $\Pi$, we conclude that $\mathcal{A}$ cannot break full anonymity of $\Theta$.
	
\end{proof}

\bibliographystyle{ieeetr}
\bibliography{OpenPubref}


\bibliographystyle{ieeetr}
\bibliography{OpenPubref}
\end{document}